\documentclass[twocolumn]{aastex701}

\usepackage{color,graphicx,natbib}
\usepackage{graphicx}
\usepackage{amsmath}
\usepackage{amssymb}
\usepackage{bm}
\usepackage[toc,page]{appendix}

\def\ra#1#2#3{#1$^{\rm h}$ #2$^{\rm m}$ #3$^{\rm s}$}
\def\dec#1#2#3{$#1^\circ #2' #3''$}

\shortauthors{Schroeder et al.}

\begin{document}

\author[0000-0001-9915-8147]
{Genevieve~Schroeder}
\affiliation{Department of Astronomy, Cornell University, Ithaca, NY 14853, USA}\email[show]{gms279@cornell.edu}

\author[0000-0002-9017-3567]{Anna~Y. Q. Ho}
\affiliation{Department of Astronomy, Cornell University, Ithaca, NY 14853, USA}\email{ayh24@cornell.edu}

\author[0009-0000-6548-6177]{Ranadeep G. Dastidar}
\affiliation{Department of Physics and Astronomy, Purdue University, 525 Northwestern Avenue, West Lafayette, IN 47907, USA}\email{rdastida@purdue.edu}

\author[0000-0001-7132-0333]{Maryam~Modjaz}
\affiliation{Department of Astronomy, University of Virginia, Charlottesville, VA 22904, USA}\email{vru7qe@virginia.edu}

\author[0000-0001-8104-3536]{Alessandra~Corsi}
\affiliation{William H. Miller III Department of Physics and Astronomy, Johns Hopkins University, Baltimore, Maryland 21218, USA}\email{acorsi2@jh.edu}

\author[0000-0001-7626-9629]{Paul C. Duffell}
\affiliation{Department of Physics and Astronomy, Purdue University, 525 Northwestern Avenue, West Lafayette, IN 47907, USA}\email{pduffell@purdue.edu}

\title{A Late-time Radio Search for Highly Off-axis Jets from PTF Broad-lined Ic Supernovae in GRB-like Host Galaxy Environments}

\begin{abstract}
    Hydrogen/Helium-poor stripped-envelope core-collapse supernovae with broad lines (SNe Ic-bl) almost always accompany the nearby ($z < 0.3$) jetted relativistic explosions known as long duration gamma-ray bursts (GRBs). However, the majority of SNe Ic-bl have no detected GRB counterpart. At least some of these SNe should harbor off-axis jets, whose afterglow may become detectable at late times, particularly at radio wavelengths.
    Here, we present Karl G. Jansky Very Large Array radio observations (rest frame times of $\sim 3$--$4\times10^{3}$ days post SN discovery) of a sample of 14 SNe Ic-bl discovered by the Palomar Transient Factory (PTF) that have been demonstrated to originate from the same host environments as the SNe Ic-bl associated with nearby GRBs. Of the 14 SNe, we identify three that are radio detected, one of which (PTF10tqv, $z = 0.0795$) is consistent with an off-axis jet with energy similar to classical GRBs (\added{${\sim 10^{51}}$--${10^{51.7}~}$erg}).
    Using recently developed synchrotron radiation code, we find that for our 11 non-detections, which are among the deepest limits obtained for Ic-bl, we rule out an off-axis jet with an energy of $\gtrsim 10^{51}~{\rm erg}$ in circumburst densities of $\gtrsim 10^{-1}~{\rm cm}^{-3}$.
    We predict that well-spaced monitoring of newly discovered SNe Ic-bl from $\sim 10~$days to $\sim 10~$years (rest frame) to luminosities of $\sim 10^{27}~{\rm erg~s}^{-1}~{\rm Hz}^{-1}$ will constrain the existence of highly off-axis jets ($\gtrsim60^\circ$) with classical GRB energies. The VLA Sky Survey will probe jets that are $\lesssim 60^\circ$ off-axis, whereas the Deep Synpotic Array 2000 will probe jets out to $\sim 90^\circ$ off-axis, demonstrating the importance of utilizing radio surveys to supplement targeted observations.
\end{abstract}

\section{Introduction}

Long duration gamma-ray bursts (GRBs) are among the most luminous explosions detected throughout the universe, with \added{beaming-corrected} energies of up to $\sim 10^{52}~$erg released on timescales of $\gtrsim 2~$s \citep{Piran_2005,Meszaros2006,Kumar_2015}. These events have been observationally linked to the death of massive stars, with their association with supernovae (SNe) and locations in star forming galaxies \citep[for reviews, see e.g.][]{WoosleyBloom2006,Fruchter2006,Cano2017, 2019NatAs...3..717M}.

Nearly all the SNe observed associated with GRBs have been spectroscopically classified as hydrogen and helium-poor stripped envelope core-collapse SNe with broad lines  (Type Ic-bl; \added{ \citealt{WoosleyBloom2006, Modjaz2016,Cano2017}}). However, there exists a large population of SNe Ic-bl  with no detected GRB counterparts \citep{2016ApJ...830...42C, 2019A&A...621A..71T,Ho2020,2023ApJ...953..179C,Srinivasaragavan2024}. One explanation for the SNe Ic-bl lacking GRBs is that all or some produce a relativistic jet, but the jet is oriented off-axis from our viewing angle ($\theta_{\rm obs}$). The existence of off-axis jets in SNe Ic-bl is a basic prediction of the model of GRBs as collimated explosions.
This theory can be tested by observing SNe Ic-bl at late times, at which point the GRB jet should have spread such that the jet opening angle ($\theta_{\rm j}$) is larger than $\theta_{\rm obs}$ \citep{Rhoads1997,Sari1999,Granot2002} and/or the relativistic beaming angle become larger than $\theta_{\rm obs}$ \citep{2010ApJ...722..235V, 2011ApJ...733L..37V}, resulting in the synchrotron afterglow from the jets becoming observable.

Radio searches for the off-axis afterglow are particularly effective because in the optical band, for a wide range of viewing angles and observing epochs, the SN emission is expected to outshine the afterglow \citep{2016MNRAS.461.1568K}. Numerous studies have utilized radio observations to search for the signature of an off-axis afterglow following \added{core-collapse} SNe \citep{2003ApJ...599..408B, 2006ApJ...638..930S, 2014MNRAS.440..821B, 2014ApJ...782...42C,  2016ApJ...830...42C, 2023ApJ...953..179C, 2019ApJ...879...89M, 2019ApJ...872..201P, 2021ApJ...910...16P}. However, these previous studies have several limitations, including small numbers of SNe Ic-bl, insufficiently sensitive radio observations, and limited temporal coverage. As a result, to date no conclusive off-axis jet has been detected following a SN Ic-bl, although there have been candidates. For example, SN\,2020bvc had early high-velocity features in its optical spectra, as well as fading X-ray emission, that was argued to be consistent with an off-axis GRB cocoon and afterglow \citep{Izzo2020}. However, the multiwavelength behavior is very similar to that of GRB\,060218/SN\,2006aj \citep{2020ApJ...902...86H}, which is widely modeled as a quasi-spherical outflow, or alternatively a jet of significantly lower power than classical\footnote{We use the term ``classical GRB\added{s}'' to describe the highly collimated, energetic events found at cosmological distances, as opposed to low-luminosity GRBs, which are thought to be quasi-spherical and may be more common \added{\citep{2004Natur.430..648S,2007ApJ...662.1111L,Cano2017}}.}  GRBs (see \citealt{Irwin2016} for a discussion of possible models). Additionally, the radio transient FIRST J141918.9+394036 is consistent with an off-axis GRB origin, however there is no associated SN Ic-bl to confirm this progenitor \citep{2018ApJ...866L..22L, 2022ApJ...924...16M}. Here we seek to expand upon previous work by observing a sample of SNe Ic-bl demonstrated to originate from GRB-like host-galaxy environments out to later times and deeper luminosities than previously achieved.

We utilize a sample of 14 SNe Ic-bl detected by the Palomar Transient Factory \citep[PTF, ][]{2009PASP..121.1395L} a galaxy-untargeted survey between 2009 and 2012. This sample is identical to the SN Ic-bl sample presented in  \citet{2020ApJ...892..153M}, who found that the hosts of these SNe were indistinguishable from a sample of GRB-SNe host galaxies, but very different from those of the environments of PTF SNe Ic. In particular, \citet{2020ApJ...892..153M} found that SNe Ic-bl and GRB-SN originate in similarly low metallicity environments, indicating that these classes of SNe are produced by similar massive star progenitors. \citet{2020ApJ...892..153M} predict that this sample of SNe Ic-bl could be off-axis analogs to on-axis GRB-SN, or may have produced jets choked inside the star.
This prediction is corroborated by theoretical modeling by \citet{2018ApJ...860...38B}, who demonstrated that jetted central engines could produce both the GRB, and at the same time, the broad-lined spectra of their accompanying SNe, even for off-axis jet cases. As a result, we have chosen this sample to test these predictions. 

In Section~\ref{sec:Observations}, we present targeted observations of our sample using the NSF's Karl G. Jansky Very Large Array (VLA), as well as archival radio observations when applicable. In Section~\ref{sec:Analysis}, we present our off-axis radio afterglow model light curves and discuss how our radio observations compare to these models. In Section~\ref{sec:Discussion} we place our results into the larger context of SNe Ic-bl and GRBs, discuss the limitations of this study, and present a path forward to  
detecting off-axis jets from newly detected Ic-bl. We conclude in Section~\ref{sec:Conclusion}.
Throughout, we employ the $\Lambda$CDM cosmological parameters of H$_{0} = 68~{\rm km \, s}^{-1} \, {\rm Mpc}^{-1}$, $\Omega_{M} = 0.31$, $\Omega_{\rm \Lambda} = 0.69$ \citep{2020A&A...641A...6P}. Additionally, we use the convention $F_{\nu} \propto t^{\alpha} \nu^{\beta}$.

\begin{figure*}[!t]
\centering
\includegraphics[width=0.8\linewidth,clip=]{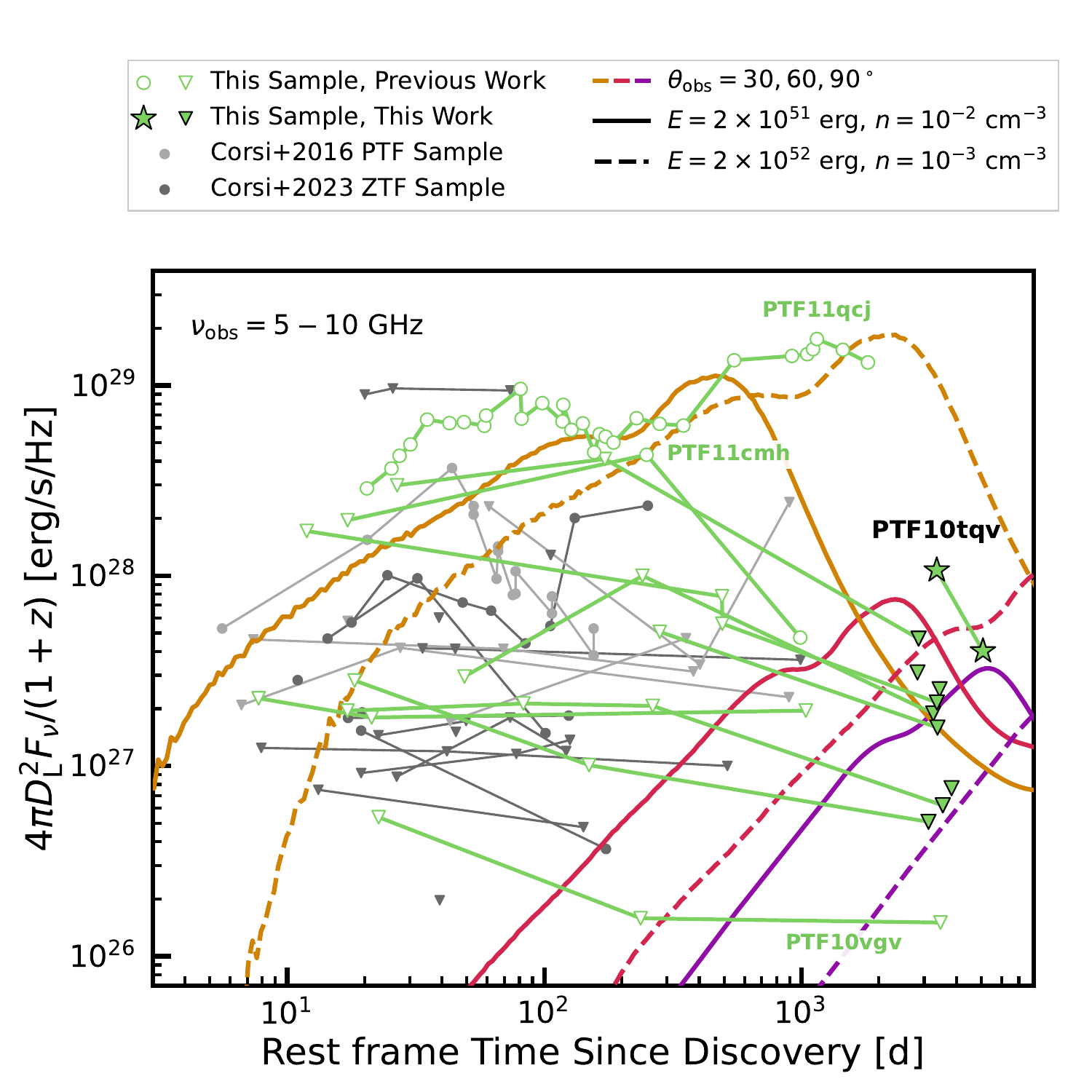} 
\caption{The 5--$10\,{\rm GHz}$ light curves and upper limits of our PTF sample (green, with observations presented in this paper filled in), compared to SNe Ic-bl samples from \citet[][PTF, light grey]{2016ApJ...830...42C} and \citet[][ZTF, dark grey]{2023ApJ...953..179C}. Also shown are off-axis 10\,GHz afterglow models generated by the \texttt{FIREFLY} code for three off-axis angles ($\theta_{\rm obs} = 30, 60, 90^\circ$; orange, pink, purple, respectively) and for two $n$-$E$ pairs ($E = 2\times 10^{51}~{\rm erg}$, $n = 10^{-2}~{\rm cm}^{-3}$, solid lines and $E = 2\times 10^{52}~{\rm erg}$, $n = 10^{-3}~{\rm cm}^{-3}$, dashed lines). Points represent detections and triangles represent upper limits ($3 \sigma$ unless otherwise stated). The detection{s} of PTF10tqv {are} represented as star{s}.
We also label two SNe Ic-bl, PTF11qcj and PTF11cmh \citep{2014ApJ...782...42C, 2016ApJ...830...42C, 2019ApJ...872..201P, 2021ApJ...910...16P}, whose luminous radio emission is thought to arise from CSM interaction, as well as the SNe Ic-bl with the deepest luminosity limits in our sample, PTF10vgv.}
\label{fig:Corsi_and_models}
\end{figure*}

\section{Observations}
\label{sec:Observations}

\begin{figure*}[!t]
\centering
\includegraphics[width=0.99\linewidth,clip=]{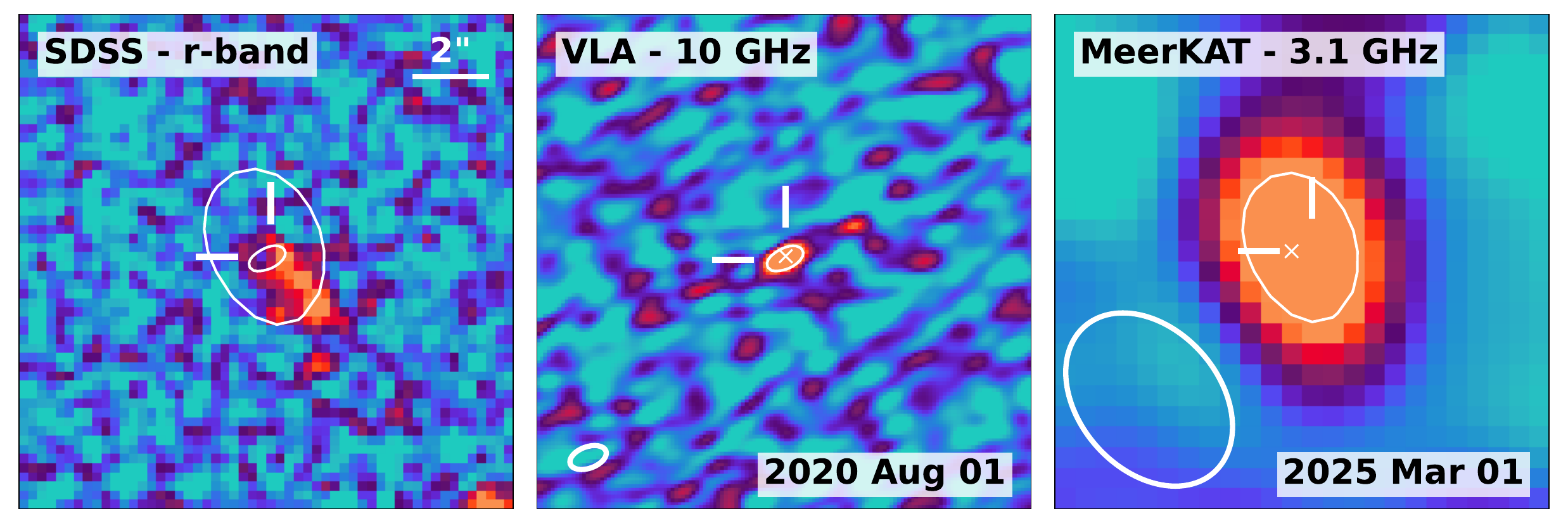} 
\caption{Left: SDSS image of the host galaxy of PTF10tqv (SDSS J224654.99+174728.5). Crosshairs represent the optical location of PTF10tqv. Middle: The 10\,GHz VLA image of the location of PTF10tqv, obtained on 2020 August 01 ($\Delta t = 3623$~days, observer frame). Right: The 3.1\,GHz MeerKAT image of the location of PTF10tqv, obtained on 2025 March 01 ($\Delta t = 5296$~days, observer frame). For the VLA and MeerKAT images, contours represent the $5\sigma$ level of the image and crosses represent the location of the brightest pixel near the location of PTF10tqv. These contours are also overplotted on the SDSS image to demonstrate the location of the radio source. The length scale is the same for all three panels, and stated in the top right of the left panel. The observing frequency is stated in the top left. When available, the beam size is represented as an ellipse in the bottom left, and the observation date is stated on the bottom right of each image.}
\label{fig:DetectionImage}
\end{figure*}

The original sample consists of 14 SNe Ic-bl detected by PTF between 2009 and 2012, which have similar host environments to local GRB-SNe \citep{2020ApJ...892..153M}.
Of the 14 SNe Ic-bl, four have extensive previous targeted radio observations, and we therefore did not pursue new radio observations, though we did search for archival radio data (Section~\ref{sec:obs_archive}). Of the four that we did not observe, two have confirmed bright radio detections: PTF11cmh and PTF11qcj \citep{2014ApJ...782...42C, 2016ApJ...830...42C}. While the origin of the radio emission for PTF11cmh remains inconclusive \citep{2016ApJ...830...42C}, continued radio monitoring of PTF11qcj, including Very Long Baseline Interferometry (VLBI), disfavors an off-axis jet origin \citep{2021ApJ...910...16P}. 

\subsection{Our Follow-up Radio Observations with the VLA and MeerKAT}
\label{sec:Radio_Observations}

We observed the remaining sample of 10 SNe Ic-bl with the VLA under program 20A-335 (PI: Ho) at a mean frequency of 10~GHz (4 GHz bandwidth) in B- and C-configuration between May 6 and September 6, 2020 ($\sim 3239$--$3970~$days post discovery). We downloaded the calibrated data set from the Science Ready Data Products\footnote{\url{https://science.nrao.edu/srdp}} initiative, which uses the Common Astronomy Software Applications \citep[\texttt{CASA}]{2007ASPC..376..127M, 2022PASP..134k4501C} VLA Calibration Pipeline\footnote{\url{https://science.nrao.edu/facilities/vla/data-processing/pipeline}}. We visually inspected and flagged the data, and then used the VLA Imaging Pipeline\footnote{\url{https://science.nrao.edu/facilities/vla/data-processing/pipeline/vipl}} to produce calibrated continuum images. To measure the flux densities and rms values at the position of each SNe, we use the \texttt{pwkit}/\texttt{imtool} program \citep{2017ascl.soft04001W}, which measures the flux density of the radio image by fitting a Gaussian within a small region (typically a width of $\sim 15$--$30~$pixels, or a $\sim$few times the size of the beam) centered on image coordinates provided by the user (here we use the coordinate of the best known position of the SN). The user can either force \texttt{imtool} to calculate the flux density of the image assuming a Gaussian the same size and shape as the telescope beam (point source), or allow the Gaussian to vary freely, in which case \texttt{imtool} can determine whether an extended profile is preferred compared to a point source based on a $\chi^2$ assessment of the fit. We choose to allow the Gaussian to vary freely in order to determine if any radio emission is extended. In the case of non-detections, we quote $3 \sigma$ upper limits equal to 3 times the rms. A summary of our observations can be found in Table~\ref{tab:radio}. We include all of our 10~GHz observations in Figure~\ref{fig:Corsi_and_models}. We briefly summarize the events that require additional attention below.

{\it PTF09sk:} The sky localization of PTF09sk is approximately $\sim 11.1 \arcmin$ offset from the quasar (and VLA flux calibrator) 3C286, which has a radio flux density of $\sim 4.5~$Jy at a mean frequency of 10~GHz \citep{2017ApJS..230....7P}.
While the primary beam field of view\footnote{\url{https://science.nrao.edu/facilities/vla/docs/manuals/oss/performance/fov}} of the VLA at 10~GHz is $\sim 4.2\arcmin$, there are significant artifacts within the image due to the proximity to 3C286.
There is a $\sim 3.5 \sigma$ source of $25.6 \pm 7.4~\mu$Jy at the location of PTF09sk in our 10~GHz image (Figure~\ref{fig:SN_VLA_Images}), but we cannot determine whether this source is real or a noise fluctuation due to the image quality. As a result, we adopt an upper limit of $\lesssim 26~\mu$Jy for this observation.

{\it PTF10tqv:} We detect a radio source coincident with the location of PTF10tqv (Figure~\ref{fig:DetectionImage}), at a position of R.A.(J2000) = \ra{22}{46}{55.042}$\pm 0.108\arcsec$ and Decl.(J2000) = \dec{+17}{47}{29.24}$\pm 0.07 \arcsec$ (the beam size of this image is $1.1\arcsec \times 0.6 \arcsec$). This source is $\sim 1\arcsec$ offset from the center of the host galaxy of PTF10tqv (SDSS J224654.99+174728.5, \citealt{2019A&A...621A..71T}), to the North East, and $\sim 0.3\arcsec$ offset from the best known position of PTF10tqv \citep{2019A&A...621A..71T}. For the radio source we measure a flux density of $F_\nu = 70 \pm 7.6~ \mu$Jy, and a point source profile is preferred by \texttt{imtool}. To measure the in-band spectral index, we bin the data into two frequency bins at 9 and 11~GHz (bin width of 2~GHz). Assuming $F_\nu \propto \nu^\beta$, we measure $\beta = -2.1 \pm 1.4$, suggesting that the emission is optically thin.

We acquired additional observations of PTF10tqv on March 1, 2025 ($\sim 5296$ days post discovery, observer frame) with MeerKAT (Project code DDT-20250212-GS-01, PI Schroeder), at a mean frequency of 3.1~GHz (bandwidth of 0.875~GHz). We download the SARAO Science Data Processor pipeline images and detect a radio source at the position of PTF10tqv (the beam size of the image is $5.5\arcsec \times 3.9 \arcsec$). We measure a flux density of $F_{\nu} = 93 \pm 12~\mu$Jy, and a point source profile is preferred by \texttt{imtool}.  
\added{We acquired additional VLA observations on September, 1 2025 ($\sim 5480$ days post discovery, observer frame) under program 25B-181 (PI Schroeder) at mean frequencies of 6 and 10~GHz (4~GHz bandwidth). We reduced and image the data following the same steps as stated above. We detect the radio source associated with PTF10tqv at both frequencies ($F_\nu = 41 \pm 12~\mu$Jy at 6~GHz and $F_\nu = 26.3 \pm 6.9~\mu$Jy at 10~GHz).}

\startlongtable
\begin{deluxetable*}{cccccccc}
\tablecolumns{8}
\tablecaption{Radio observations of the Ic-BL SNe in our sample. 
\label{tab:radio}}
\tablehead{\colhead{SN Name} & \colhead{$z$} & \colhead{Discovery Date}  & \colhead{Observed Date} &  \colhead{$\Delta t_\mathrm{obs}$} & \colhead{Observed Frequency} & \colhead{Flux Density$^{a}$} & \colhead{Observatory} 
\\ 
\colhead{} & \colhead{} & \colhead{} & \colhead{} & \colhead{(d)} & \colhead{(GHz)} & \colhead{($\mu$Jy)} & \colhead{}
}
\startdata
\hline
\multicolumn{8}{c}{Targeted Follow-up} \\
\hline
PTF09sk & 0.0355 & 2009-06-24 & 2020-05-07 & 3970 & 10.0 & $\lesssim 26^{b}$ & VLA \\
PTF10bzf & 0.0498 & 2010-03-01 & 2020-05-07 & 3720 & 10.0 & $\lesssim 10$ & VLA \\
PTF10ciw & 0.115 & 2010-03-05 & 2020-09-06 & 3838 & 10.0 & $\lesssim 7.8$ & VLA \\
PTF10qts & 0.0907 & 2010-08-08 & 2020-08-20 & 3665 & 10.0 & $\lesssim 11$ & VLA \\
PTF10tqv & 0.0795 & 2010-08-31 & 2020-08-01 & 3623 & 10.0 & $70.0 \pm 7.6$ & VLA \\
 &  &  & 2020-08-01 & 3623 & 9.0 & $82 \pm 12$ & VLA \\
 &  &  & 2020-08-01 & 3623 & 11.0 & $54 \pm 13$ & VLA \\
 &  &  & 2025-03-01 & 5296 & 3.1 & $93 \pm 12$ & MeerKAT \\
 &  &  & \added{2025-09-01} & \added{5480} & \added{6.0} & \added{$41 \pm 12$} & \added{VLA} \\
  &  &  & \added{2025-09-01} & \added{5480} & \added{10.0} & \added{$26.3 \pm 6.9$} & \added{VLA} \\
PTF10vgv & 0.015 & 2010-09-15 & 2020-05-25 & 3540 & 10.0 & $\lesssim 28$ & VLA$^{c}$ \\
PTF10xem & 0.0567 & 2010-10-03 & 2020-07-11 & 3569 & 10.0 & $\lesssim 21$ & VLA \\
PTF10aavz & 0.063 & 2010-11-17 & 2020-05-06 & 3458 & 10.0 & $\lesssim 20$ & VLA \\
PTF11gcj & 0.148 & 2011-06-21 & 2020-05-07 & 3243 & 10.0 & $\lesssim 5.7$ & VLA \\
PTF11img & 0.158 & 2011-07-22 & 2020-08-07 & 3304 & 10.0 & $\lesssim 7.5$ & VLA \\
PTF11lbm & 0.039 & 2011-08-29 & 2020-07-11 & 3239 & 10.0 & $\lesssim 14^{b}$ & VLA \\
\hline
\multicolumn{8}{c}{Radio Survey Observations} \\
\hline
PTF10ciw & 0.115 & 2010-03-05 & 2019-04-17 & 3330 & 3.0 & $\lesssim 347$ & VLASS \\
 &  &  & 2021-01-02 & 3956 & 1.367 & $\lesssim 538$ & RACS-Mid \\
 &  &  & 2021-10-01 & 4228 & 3.0 & $\lesssim 365$ & VLASS \\
 &  &  & 2022-01-07 & 4326 & 1.655 & $\lesssim 640$ & RACS-High \\
 &  &  & 2024-06-10 & 5211 & 3.0 & $\lesssim 388$ & VLASS \\
PTF11cmh & 0.1055 & 2011-04-27 & 2017-10-28 & 2376 & 3.0 & $\lesssim 235$ & VLASS \\
 &  &  & 2018-08-31 & 2683 & 0.15 & $\lesssim 132$ & LoTSS-DR2$^{d}$ \\
 &  &  & 2020-09-20 & 3434 & 3.0 & $513 \pm 157$ & VLASS \\
 &  &  & 2022-07-29 & 4111 & 1.367 & $\lesssim 507$ & RACS-Mid \\
 &  &  & 2023-02-26 & 4323 & 3.0 & $585 \pm 142$ & VLASS \\
 &  &  & 2024-03-21 & 4712 & 1.655 & $\lesssim 656$ & RACS-High \\
PTF11gcj$^{e}$ & 0.148 & 2011-06-21 & -- & -- & 0.15 & $\lesssim 232$ & LoTSS-DR2$^{d}$ \\
 &  &  & 2019-05-04 & 2874 & 3.0 & $\lesssim 871$ & VLASS \\
 &  &  & 2020-12-21 & 3471 & 1.367 & $\lesssim 1211$ & RACS-Mid \\
 &  &  & 2021-11-14 & 3799 & 3.0 & $\lesssim 702$ & VLASS \\
 &  &  & 2021-12-26 & 3841 & 1.3655 & $\lesssim 1681$ & RACS-High \\
 &  &  & 2024-06-12 & 4740 & 3.0 & $\lesssim 636$ & VLASS \\
PTF11qcj & 0.028 & 2011-11-02 & 2014-07-14 & 985 & 0.15 & $1890 \pm 155$$^{f}$ & LoTSS-DR2$^{d}$ \\
 &  &  & 2020-12-20 & 3336 & 1.367 & $5812 \pm 234$ & RACS-Mid \\
 &  &  & 2021-11-22 & 3673 & 3.0 & $3577 \pm 210$ & VLASS \\
 &  &  & 2021-12-25 & 3706 & 1.655 & $5406 \pm 244$ & RACS-High \\
 &  &  & 2024-06-29 & 4623 & 3.0 & $1807 \pm 171$ & VLASS \\
PTF12as & 0.0332 & 2012-01-02 & 2018-01-06 & 2196 & 3.0 & $\lesssim 467$ & VLASS \\
 &  &  & 2020-09-12 & 3176 & 3.0 & $\lesssim 375$ & VLASS \\
 &  &  & 2021-01-09 & 3295 & 1.367 & $\lesssim 440$ & RACS-Mid \\
 &  &  & 2022-01-07 & 3658 & 1.655 & $\lesssim 551$ & RACS-High \\
 &  &  & 2023-02-07 & 4054 & 3.0 & $\lesssim 339$ & VLASS \\\hline
\enddata
\tablenotetext{a}{Uncertainties correspond to $1\sigma$ confidence. Upper limits correspond to $3 \sigma$.}
\tablenotetext{b}{A $\lesssim 5 \sigma$ source is found consistent at or near the position of the SN, however it is not possible to confirm the realness of the source and therefore we quote a $3 \sigma$ upper limit instead See Section~\ref{sec:Radio_Observations}).}
\tablenotetext{c}{Observations were taken under Program 20A-491 (PI Stroh, See Section~\ref{sec:obs_archive}).} 
\tablenotetext{d}{The LoTSS-DR2 cutout images are derived from mosaic images which may be formed from multiple observations. For PTF11cmh and PTF11qcj, the sky position is only covered by a single pointing, and therefore we quote the observation date of the pointing. For PTF11gcj, the position is covered by two pointings observed on 2016 Mar 24 and 2019 Jun 06, as a result we do not quote an observation date.}
\tablenotetext{e}{The position of PTF11gcj is $\sim 20.2\arcmin$ from the radio quasar 3C345, resulting in high rms values (See Section~\ref{sec:obs_archive}).} 
\tablenotetext{f}{The radio source at the position of PTF11qcj in the LoTSS-DR2 image is extended, and therefore it is unlikely that all of the emission detected originates from the SN (See Section~\ref{sec:obs_archive}).}
\end{deluxetable*}

{\it PTF10xem:} We detect a radio source at R.A.(J2000) = \ra{01}{47}{07.028}$\pm 0.068\arcsec$ and Decl.(J2000) = \dec{+13}{56}{29.15}$\pm 0.05 \arcsec$ (Figure~\ref{fig:SN_VLA_Images}; the beam size of the image is $1.0\arcsec \times 0.6 \arcsec$), coincident with the position of the host galaxy of PTF10xem, SHOC 084 \citep{2019A&A...621A..71T}. For the radio source, we measure a flux density of $F_\nu = 135 \pm 10~\mu$Jy, and a point source profile is preferred. If we assume this radio emission can be attributed solely to star formation, this would imply a radio derived star formation rate of ${\rm SFR}_{\rm radio} = 2.99 \pm 0.23~M_\odot/{\rm yr}$ (following the prescription of \citet{2016A&A...593A..17G}), similar to the optically derived SFR of $2.63^{+0.32}_{-0.06}~M_\odot/{\rm yr}$ from \citet{2020ApJ...892..153M}. We do not detect any radio emission at the position of PTF10xem, which was localized to $\sim 2.5\arcsec$ from the center of the host galaxy. Given that the beam size of the image is smaller than the SN offset from the host galaxy, we conclude that the radio emission we detect is unrelated to the SN.

{\it PTF11lbm:} There is a $\sim 2.8 \sigma$ source at R.A.(J2000) = \ra{23}{48}{03.195}$\pm 0.213\arcsec$ and Decl.(J2000) = \dec{+26}{44}{34.05}$\pm 0.203 \arcsec$, and we measure a flux density of $14.1 \pm 5.1~\mu$Jy (Figure~\ref{fig:SN_VLA_Images}, the beam size of the image is $0.6\arcsec \times 0.6\arcsec$). This source is $\sim 0.6\arcsec$ from the published location of PTF11lbm \citep{2016ApJ...830...42C, 2019A&A...621A..71T}. With our current data, it is not possible to determine whether this source is real or a noise fluctuation, and as a result, we adopt an upper limit of $\lesssim 14~\mu$Jy for this observation. 

\subsection{Archival Observations}
\label{sec:obs_archive}

We performed an archival search for radio observations at the locations of our 14 SNe. This search includes the VLA archive\footnote{\url{https://data.nrao.edu/portal/}}, the VLA Sky Survey\footnote{We retrieve cutouts from the 
Canadian Initiative for Radio Astronomy Data Analysis (CIRADA) Image Cutout Web Service (\url{cutouts.cirada.ca/})} at a mid-frequency of 3~GHz \citep{2020PASP..132c5001L},  LOw-Frequency ARray (LOFAR)
Two-metre Sky Survey (LoTSS) Data Release 2 (DR2\footnote{We generate cutouts using the LoTSS DR2 Cutout Service (\url{https://lofar-surveys.org/dr2\_release.html})}) at a mid-frequency of $150~$MHz \citep{2022A&A...659A...1S}, and the Australian SKA Pathfinder (ASKAP) Rapid ASKAP Continuum Survey (RACS\footnote{We retrieve cutouts from the Commonwealth Scientific and Industrial Research Organisation (CSIRO) ASKAP Science Data Archive (CASDA; \url{https://data.csiro.au/domain/casdaCutoutService})}, \citealt{2020PASA...37...48M}) at mid-frequencies of 1.37~GHz \citep[RACS-Mid, ][]{2023PASA...40...34D, 2024PASA...41....3D} and 1.66~GHz \citep[RACS-High, ][]{2025arXiv250104978D}. Furthermore, we include data published in the literature \citep{2014ApJ...782...42C, 2016ApJ...830...42C, 2019ApJ...872..201P, 2021ApJ...910...16P}.

For the majority of the events, we do not detect any significant emission at or near the position of the SN in any of the survey images. Given the typical depths of the survey images (rms of $\sim 120~\mu$Jy for VLASS\footnote{\url{https://science.nrao.edu/science/surveys/vlass}}, $\sim 200~\mu$Jy for RACS-Mid and RACS-High\footnote{\url{https://research.csiro.au/racs/home/survey/}}, and $\sim 83~\mu$Jy for LoTSS-DR2\footnote{\url{https://lofar-surveys.org/dr2\_release.html}}), the non-detections are usually not more constraining than targeted VLA observations. As such, we only consider the non-detections for two Ic-bl with no other targeted radio observations (PTF10ciw and PTF11gcj; we omit PTF09sk as the image artifacts from 3C286 persist in the survey images), as well as PTF12as, which was only observed out to $\sim 1074~$days (observer frame) with prior radio campaigns \citep{2016ApJ...830...42C}. Additionally, we include survey data for two events with prior radio detections: PTF11cmh and PTF11qcj (see below, \citealt{2014ApJ...782...42C, 2016ApJ...830...42C, 2019ApJ...872..201P, 2021ApJ...910...16P, 2021ApJ...923L..24S}). We summarize these observations in Table~\ref{tab:radio}, and discuss events that require additional attention below.

{\it PTF10tqv:} The field of PTF10tqv was observed three times (20 May 2019, 6 Nov 2021, and 4 July 2024) by VLASS; however, the rms of the survey images ($\sim 100$--$130~\mu$Jy) prevent us from determining if the radio source we detect in our 3.1 GHz MeerKAT images is present in VLASS. 

{\it PTF10vgv:} Observations were obtained using the VLA on 2020 May 25 at a mean frequency of 10~GHz (4~GHz bandwidth) under Program 20A-491 (PI Stroh). We follow the same steps laid out in Section~\ref{sec:Radio_Observations} to download and image the data. We do not detect any emission at the location of PTF10vgv, to a $3 \sigma$ limit of $\lesssim 30~\mu$Jy. 

{\it PTF11cmh:} PTF11cmh has previously been detected by the VLA out to $\sim 1136$~days at 2.9~GHz \citep{2016ApJ...830...42C}. The position of PTF11cmh was observed by VLASS on 2017 Oct 28 (Epoch 1.1), 2020 Sep 20 (Epoch 2.1), and 2023 Feb 26 (Epoch 3.1). No emission is detected at the position of PTF11cmh in the Epoch 1.1 image, to a $3\sigma$ limit of $\lesssim 235~\mu$Jy. However, in Epoch 2.1 and 3.1, there is a $\sim 3$--$4\sigma$ source near the position of PTF11cmh (R.A. (J2000) = \ra{13}{10}{21.7} Decl. (J2000) = \dec{+37}{53}{02} in Epoch 2.1 and R.A. (J2000) = \ra{13}{10}{21.7} Decl. (J2000) = \dec{+37}{52}{59} in Epoch 3.1, with a positional uncertainty of 1\arcsec), and we measure a flux density of $F_\nu \approx 513$--$585~\mu$Jy.
We note that the position of the source in VLASS Epoch 2.1 is not fully consistent with the position of PTF11cmh, and is offset by $\sim 2\arcsec$, whereas the position of the source in Epoch 3.1 is only offset by $\sim 0.6\arcsec$. Given the low significance of the source, we can not confidently claim a detection.
No emission is detected at or near the position of PTF11cmh in LoTSS-DR2 or RACS. These images are presented in Figure~\ref{fig:DetectionImage_PTF11cmh}. We explore this possible detection more in Section~\ref{sec:PTF11cmh}.

{\it PTF11gcj:} The position of PTF11gcj is $\sim 20.2\arcmin$ away from the radio quasar 3C345, which is $\sim$few Jy at radio wavelengths \citep{1986ApJ...301..708B}. While this source does not appear to affect the image quality of our VLA observation, we note some of the survey images appear to have significant artifacts at/near the location of PTF11gcj, and the image rms is higher than expected for the majority of the survey images.

{\it PTF11qcj:} PTF11qcj has previously been detected at radio wavelengths out to $\approx 2829$--$2857$~days at 1.5--5.1~GHz \citep{2014ApJ...782...42C, 2019ApJ...872..201P, 2021ApJ...910...16P}, as well as in VLASS \citep[2019 March 19, Epoch 1.2]{2021ApJ...923L..24S}. We further confirm that PTF11qcj is detected in VLASS on 2021 Nov 22 and 2024 June 29 (Epoch 2.2 and 3.2, respectively). The measured flux density from VLASS declines from $\sim 6.8~$mJy in Epoch 1.2 \citep[$\sim 2695~$days since discovery, ][]{2021ApJ...923L..24S} to $\sim 1.8~$mJy in Epoch 3.2 ($\sim 4624~$days since discovery), and indicates fading from the last targeted VLA observations at a similar frequency \citep[$\sim 1863~$days since discovery,][]{2019ApJ...872..201P}.

The position of PTF11qcj was observed by RACS-Mid, RACS-High, and LoTSS-DR2 (Figure~\ref{fig:DetectionImage_PTF11qcj}), and a source at the position of PTF11qcj is detected in all images. However, in the LoTSS-DR2, \texttt{imtool} prefers an extended profile for the source (approximate size of $13.8\arcsec \times 9.2\arcsec$, compared to a beam size of $6\arcsec \times 6\arcsec$), indicating the radio emission may originate from the host galaxy \citep[LEDA 2295826, ][]{2021ApJ...908...75B} rather than the SN itself. We measure a flux density of $\sim 1.9~$mJy at $\sim 985~$days post discovery. Targeted VLA observations from $2.5$--$16~$GHz were obtained at a similar epoch \citep[$\sim 967~$days, ][]{2019ApJ...872..201P}. The VLA data are well fit by a broken power law with a peak of $\sim 8~$mJy at $\sim 6~$GHz  and spectral indices of $\beta_1 \approx 1.1$ and $\beta_{2} \approx -0.9$. Extrapolating from this broken power law, we find the expected $150~$MHz flux density to be $\sim 180~\mu$Jy for this epoch. This is $\sim 10 \times$ lower than our measured value, further indicating that the detected emission is unrelated to the SN and rather is due to host galaxy emission.
If the emission detected in the LoTSS-DR2 image can be attributed solely to star formation, this would imply ${\rm SFR}_{\rm radio} = 0.42 \pm 0.03~M_\odot/{\rm yr}$, 3 times less than the optically derived SFR of $1.35_{-0.33}^{+0.56}~M_\odot/{\rm yr}$ \citep{2020ApJ...892..153M}. 

The RACS-Mid and RACS-High source is similar in size to the beam ($39.8\arcsec \times 11.3 \arcsec$ and $36.8\arcsec \times 6.1 \arcsec$, respectively), and therefore it is not possible to distinguish whether the emission is from the SN or host galaxy based on the source shape alone. However, if the RACS images were significantly contaminated by host emission, we would expect the radio brightness to be significantly higher than the 1.5~GHz VLBI measurement \citep[$\sim 5.8~$mJy, taken 2019 Aug 29,][]{2021ApJ...910...16P}, as the host emission should be resolved out in the VLBI image. Instead, the flux density measurements from the RACS images ($\sim 5.8~$mJy at 1.4~GHz and $\sim 5.4~$mJy at 1.7~GHz, taken on 2020 Dec 20 and 2021 Dec 25, respectively) are similar to the VLBI radio brightness at 1.5 GHz, indicating that there is little contamination from the host galaxy in the RACS images.

\section{Analysis}
\label{sec:Analysis}

Here, we place our observations into the context of radio afterglows from off-axis GRBs. We first describe how we generate off-axis light curves, and then use these generated light curves to constrain the presence of an off-axis jet for the SNe Ic-bl in our sample in order to connect our data to the physical properties of GRBs.

\begin{figure*}[!t]
\centering
\includegraphics[width=0.99\linewidth,clip=]{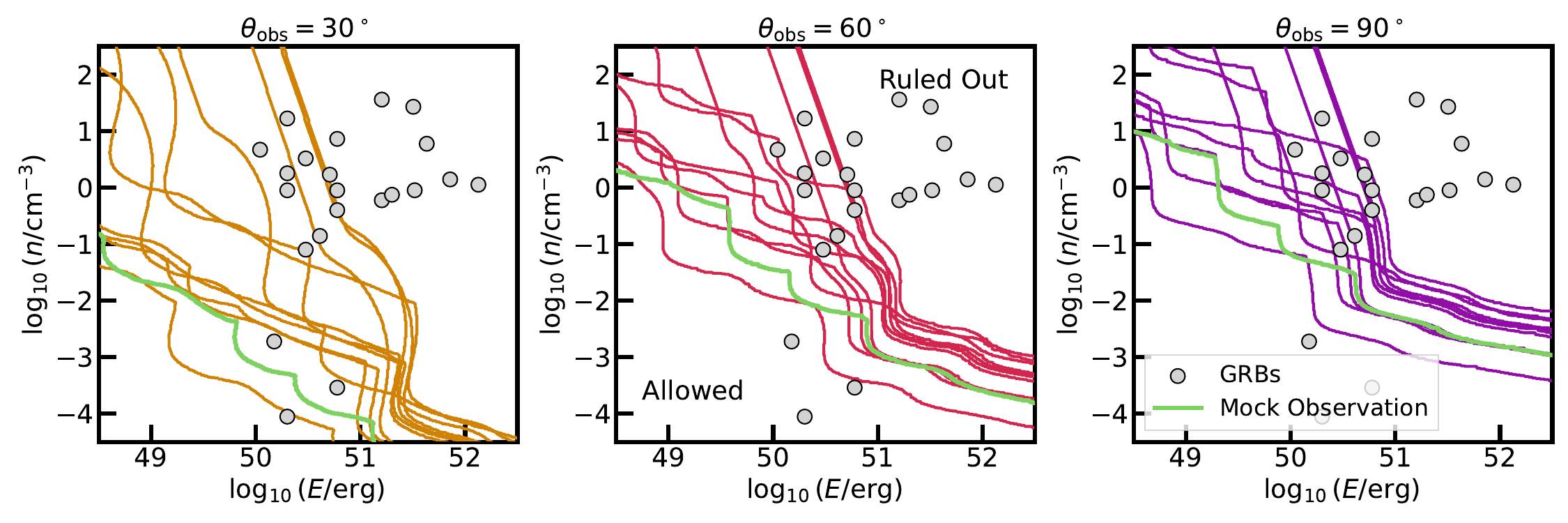} 
\caption{The interstellar medium density ($n$) vs jet energy ($E$) parameter space for three off-axis observer angles ($\theta_{\rm obs} = 30, 60, 90^\circ$), from models generated by the \texttt{FIREFLY} code. Lines represent the constraints on the parameter space from each of our 11 non-detections, where the upper right of each line is ruled out and the lower left of each line is still allowed. We similarly display the constraints that can be placed on the $n$-$E$ parameter space given our suggested observational strategy of future SNe Ic-bl (Section~\ref{sec:Discussion}, green).
Also shown are the beaming corrected kinetic energy and $n$ values for some on-axis GRBs \citep{2002ApJ...571..779P, 2021ApJ...911...14K}.}
\label{fig:Energy_vs_Density}
\end{figure*}

\subsection{Model Light Curves}
\label{sec:Models}

We assume a relativistic jet expanding into a constant-density interstellar medium\footnote{\added{While a wind-like density ($\rho \propto r^{-2}$, where $r$ is the radius) is physically expected for the massive star progenitors of long GRBs, we assume a constant-density interstellar medium given the large implied radius of a GRB-like jet at the time of our observations ($\sim 10^{19}~$cm at $\sim 10~$years).}} to model the off-axis GRB afterglow. We consider a two-parameter boosted fireball to model the jet \citep{Duffell+2013}. 
In this model, the jet is launched as a fireball with a bulk Lorentz factor ($\eta_0 \sim E/M$) and boosted in a particular direction with a Lorentz factor, $\gamma_{\rm{B}}$. 
The Lorentz boost sets the jet opening angle, $\theta_{\rm j} \sim 1/\gamma_{\rm B}$. 
In this study, we use $\eta_0 = 6.5$ and $\gamma_{\rm B} = 8.0$ which gives $\theta_{\rm j} \sim 7^{\circ}$ (similar jet opening angles for core-collapse SNe have been obtained by \citealp{Mizuta+2013, Duffell+2015}). However, at later times (more relevant to this study), the jet spreads out, and the initial jet opening angle becomes less relevant. The model afterglow light curves are then generated using the \texttt{FIREFLY} code \citep{2024ApJ...976..252D}. 
The \texttt{FIREFLY} code is based on standard GRB afterglow theory \citep{1998ApJ...497L..17S} and openly accessible\footnote{\url{https://doi.org/10.5281/zenodo.15699685}}. 
The \texttt{FIREFLY} code has been tested against \texttt{BOXFIT} \citep{2012ApJ...749...44V} and the light curves produced by both codes are in agreement with each other.  
We note that our models do not include any contribution from the supernova to the overall radio emission, though other models have included this contribution \citep[e.g.][]{2015MNRAS.454.1711B, 2016MNRAS.461.1568K,2025arXiv250313291K}. Our models also do not take synchrotron self absorption into account, which is expected to affect low radio frequencies ($\lesssim 1~$GHz).

Consequently, our code has six free parameters to model the \added{light curves}: the total jet energy ($E$, beaming corrected), the interstellar medium density ($n$), the power-law index $p$ for the electron distribution in the shock front ($n \propto \gamma^{-p}$), the kinetic and magnetic energy fraction to the total internal energy $\epsilon_{\rm e}$ and $\epsilon_{\rm B}$, respectively, and the fraction of electrons contributing to the afterglow $\chi$. Variations in $\chi$ do not significantly change our light curves, so we fix $\chi = 1$. We additionally fix $p = 2.133$, and $\epsilon_{\rm e} = \epsilon_{\rm B} = 0.1$. Overall, these values are similar to what is measured in broad-band afterglow modeling of on-axis long GRBs \citep[e.g.][]{2002ApJ...571..779P, 2003ApJ...597..459Y, 2010ApJ...711..641C, 2011ApJ...732...29C, 2014ApJ...781....1L,  2015ApJ...799....3R, 2021ApJ...911...14K, 2022ApJ...940...53S}, though we note that $\epsilon_{\rm B} \ll 0.1$ is measured in some of these studies. We return to this point in Section~\ref{sec:Discussion}. For a given set of parameters, one can generate light curves at any user input observer angle.

To compare our observations to our off-axis afterglow models, we generate a grid of model \added{light curves}, varying $E$ and $n$ for three off-axis viewing angles, $\theta_{\rm obs} = 30^\circ, 60^\circ, 90^\circ$ and three observing frequencies, $\nu_{\rm obs} = 3, 6, 10~$GHz, to best match the observations presented in Section~\ref{sec:Observations}. 
Thus, a number of light curves $F_\nu(t)$ are generated for a grid of jet energies $\{E\}$ and \added{interstellar medium} densities $\{n\}$.  Combining the model light curves with an observed non-detection of an afterglow signal at a given frequency and observer time provides an excluded parameter space of $(n,E)$ at that observer angle. Similarly, this treatment can be extended to all observer angles.  Additionally, scale invariance of the synchrotron model implies that the light curves can be generated by a simple scaling relation, from a given fiducial \added{light curve}. We will explore these scaling relations in detail in an upcoming paper (Dastidar et al., in preparation).

\subsection{Non-detections: Constraints on Off-axis Jets}
\label{sec:Analysis_nondet}

Our 10~GHz radio non-detections probe luminosities of $\sim 1.5\times 10^{26}$--$4.7 \times 10^{27}~{\rm erg~s}^{-1}{\rm ~Hz}^{-1}$ at rest frame times of $\sim 2820$--$3830~{\rm days}$ (Figure~\ref{fig:Corsi_and_models}). The large range of luminosities is owed mostly to the redshift range of the SNe in our sample ($z = 0.015$--$0.158$), and our deepest luminosity constraints are from PTF10vgv ($z = 0.015$, $\lesssim 1.5 \times 10^{26}~{\rm erg~s}^{-1}{\rm ~Hz}^{-1}$). 
Our observations presented here of the 
\citet{2020ApJ...892..153M} PTF sample extend to later times compared to previous radio observations of SNe Ic-bl \citep[e.g.][]{2012ApJ...747L...5C, 2016ApJ...830...42C, 2023ApJ...953..179C, 2019ApJ...879...89M}.

We use all $3$--$10~$GHz observations of our Ic-bl sample, as well as the $E$ and $n$ scaling relations described in Section~\ref{sec:Models}, to determine the constraints placed on the $n$-$E$ parameter space for the three $\theta_{\rm obs}$ considered. To do so, we use the grid of generated models to determine what $n$-$E$ pair corresponds to the upper limit on the luminosity for each observation at the particular observing frequency. We plot the constraints for the 11 non-detections in Figure~\ref{fig:Energy_vs_Density}, where the parameter space to the upper right of the lines is ruled out by our non-detections and the parameter space to the lower left of each line is still consistent with our non-detections. For comparison, we also plot a sample of $n$-$E$  parameter pairs measured from long GRB afterglows \citep{2002ApJ...571..779P, 2021ApJ...911...14K}. 

For the $n$-$E$ pairs derived from long GRBs, the measured $E$ spans two orders of magnitude ($\sim 10^{50}$--$10^{52}~{\rm erg}$), whereas the measured $n$ spans six orders of magnitude $\sim 10^{-4}$--$10^2~{\rm cm}^{-3}$ \citep{2002ApJ...571..779P, 2021ApJ...911...14K}. We are able to rule out the majority of measured long GRB $n$-$E$ pairs for most of our Ic-bl sample, especially at $E > 10^{51}~{\rm erg}$ \citep[the typical energy of a GRB;][]{2001ApJ...562L..55F, 2003ApJ...594..674B, 2014PASA...31....8G, 2018IJMPD..2742002V}, for all $\theta_{\rm obs}$. However, at $\theta_{\rm obs} \gtrsim 60^\circ$, we cannot rule out the measured long GRB $n$-$E$ pairs at $n < 10^{-2}~{\rm cm}^{-2}$. 

Overall, the radio non-detections of our sample are able to rule out off-axis models of $\lesssim 30^\circ$ for the majority of the parameter space considered. We note that the low-$E$ ($\lesssim 10^{50.5}~{\rm erg}$) and high-$n$ ($\gtrsim 10^{0}~{\rm cm}^{-3}$) parameter space is difficult to constrain for SNe with no early ($\sim 10$--$100~$days) observations, as the off-axis light curves for that portion of parameter space usually peak earlier than $\sim 3000~{\rm days}$, the typical time of our observations. We return to this point in Section~\ref{sec:Discussion}.

While the additional constraints placed on the $n$-$E$ parameter space by our observations vary for each SN, typically at $E \gtrsim10^{51}~{\rm erg}$, our \added{late-time} observations are able to constrain $n$ by $\sim 1$--$2$ orders of magnitude more than the observations of the sample presented in \citet{2016ApJ...830...42C}. This is particularly important for $\theta_{\rm obs} \gtrsim 60^\circ$ models, where without these \added{late-time} observations, the parameter space at $E \gtrsim 10^{51}~{\rm erg}$ and $n \lesssim 10^{-1}~{\rm cm}^{-3}$ would remain relatively unconstrained.
Therefore, these later observations allow for constraints to be placed on late rising radio emission from highly off-axis ($\theta_{\rm obs} \gtrsim 60^\circ$) jets. 

In addition to an off-axis GRB origin for this PTF Ic-bl sample, \citet{2020ApJ...892..153M} suggested that these SNe may have produced jets that did not break out of the star (choked, \citealt{2001PhRvL..87q1102M, 2012ApJ...749..110B}) or were able to break out of the star but were unseen low-luminosity GRBs, similar to GRB\,060218/SN\,2006aj \citep{2006ApJ...638..930S}, though we note GRB\,060218/SN\,2006aj may have itself been a choked jet \citep[e.g.][]{2006Natur.442.1008C, 2007ApJ...667..351W,2015ApJ...807..172N, 2024arXiv241206736I}. The radio light curve of GRB\,060218/SN\,2006aj fades very quickly ($\lesssim 10^{27}~{\rm erg~s}^{-1}~{\rm Hz}^{-1}$ by a rest frame time of $\sim 20~{\rm days}$ post discovery), and our \added{late-time} radio observations presented here are not suited to probe such a scenario. We discuss how we could could improve upon our observational strategy in Section~\ref{sec:Discussion}.

\subsection{PTF11cmh}
\label{sec:PTF11cmh}
We now explore the possible detection of PTF11cmh in VLASS images, although as discussed in Section~\ref{sec:obs_archive}, the association is not confirmed. The last known radio detection of PTF11cmh at $\sim 3$~GHz was $18 \pm 4~\mu$Jy at $1136~$days after optical discovery \citep{2016ApJ...830...42C}. Therefore, if the VLASS detections are real, they would imply a rise rate of $\alpha \approx 3$ and a luminosity of $\sim 10^{29}~{\rm erg~s}^{-1}~{\rm Hz}^{-1}$. This behavior is not consistent with either the CSM interaction or the off-axis model presented in \citet{2016ApJ...830...42C}. For our off-axis models, a jet energy of $E \gtrsim 10^{53}~{\rm erg}$ is required to achieve such a late rise and high luminosity, $\gtrsim 2$ orders of magnitude higher than a typical GRB.

\begin{figure*}[!t]
\centering
\includegraphics[width=0.99\linewidth,clip=]{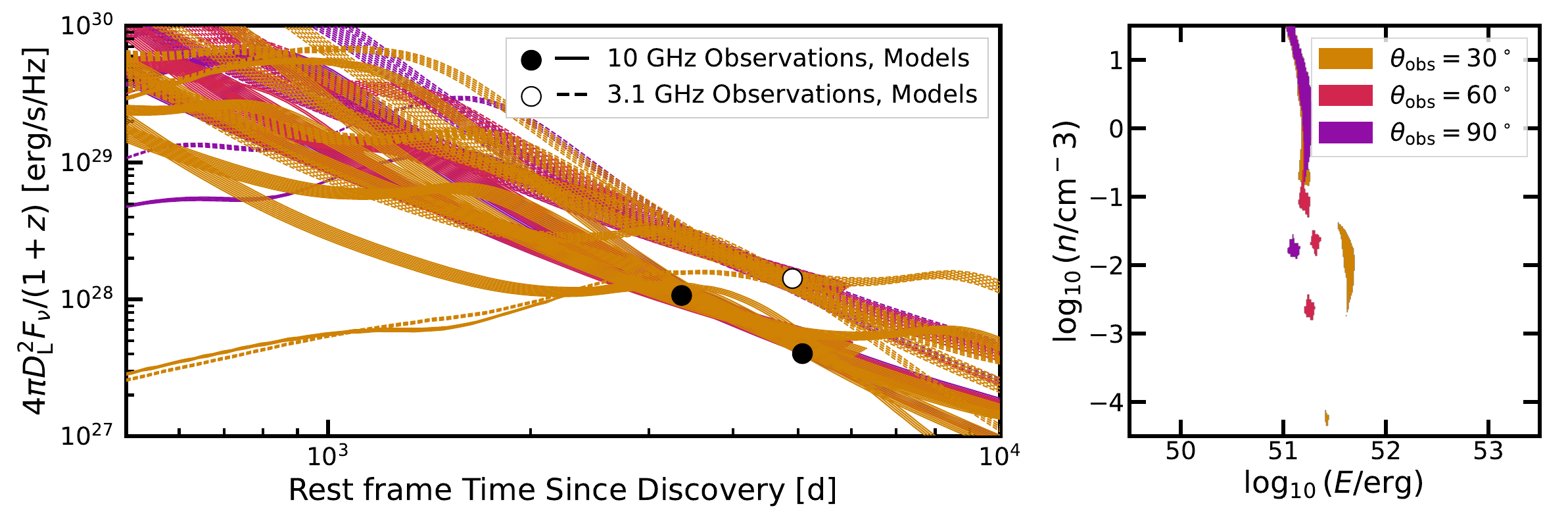} 
\caption{\emph{Left:} The 10~GHz (black) and 3.1~GHz (white) radio detections of PTF10tqv. Also shown are off-axis afterglow models consistent within $2\sigma$ of the detections, for off-axis angles of $\theta_{\rm obs} = 30, 60, 90^\circ$ (purple, orange, pink, respectively), where solid lines represent 10~GHz models and dashed lines represent 3~GHz models. \emph{Right: } The circumburst density ($n$) vs total jet energy ($E$) space for the off-axis models generated by \texttt{FIREFLY}. Shaded regions represent the space consistent within $2 \sigma$ of the radio detections of the counterpart to PTF10tqv.}
\label{fig:Energy_vs_Density_PTF10tqv}
\end{figure*}

Our off-axis models also predict an optically thin radio spectral index of $\beta \approx -0.5$ at late times, which we can test with the archival radio observations. The RACS-Mid non-detection ($\lesssim 507~\mu$Jy at a rest frame time of $\sim 3720~$days) is nearly concurrent with the VLASS Epoch 3 detection ($\sim 585~\mu$Jy at a rest frame time of $\sim 3910~$days). Therefore, we can use these observations to calculate the 1.4~GHz to 3~GHz spectral index, and we find $\beta \gtrsim 0.2$, indicating that the radio emission must be optically thick. 
Given the large required $E$ and the inconsistent spectral index, this apparent rebrightening, if real, is unlikely to be the result of an off-axis jet. Future radio observations confirming the rebrightening and spectral index will be necessary to test this assumption. For the purposes of this work we do not consider PTF11cmh as candidate off-axis GRB.

\subsection{PTF11qcj}
Radio observations of PTF11qcj have revealed luminous, multi-peaked radio emission (Figure~\ref{fig:Corsi_and_models}). Modeling of the radio emission in addition to VLBI observations of PTF11qcj have demonstrated that an off-axis jet origin for the radio behavior is unlikely \citep{2014ApJ...782...42C, 2019ApJ...872..201P, 2021ApJ...910...16P}. As a result, we do not consider PTF11qcj as a candidate off-axis GRB.

\subsection{PTF10tqv}
\label{sec:PTF10tqv}

Finally, we explore the possible sources of emission for the radio detection of PTF10tqv.  We note that all of our non-detections at 10~GHz are below the luminosity of the detection of PTF10tqv, despite its intermediate redshift ($z = 0.0795$, Figure~\ref{fig:Corsi_and_models}). Therefore, we are able to rule out PTF10tqv-like emission for the 10 other SNe Ic-bl with 10~GHz observations at rest frame times of $\sim 3$--$4 \times 10^{3}~{\rm days}$, demonstrating that it is unique in its radio detection among our sample.

\subsubsection{Radio Emission from Host Galaxy}
\added{We first explore whether the radio emission associated with PTF10tqv could be attributed to the host galaxy.}
As mentioned in Section~\ref{sec:Radio_Observations}, the 10~GHz counterpart to PTF10tqv is $\sim 1\arcsec$ from the center of the host galaxy SDSS J224654.99+174728 (Figure~\ref{fig:DetectionImage}), corresponding to $\sim 2~$kpc at a redshift of $z = 0.0795$. \added{The spectral index between the 3.1~GHz and 10~GHz detections at $\sim 5296$--$5480~$days post discovery is $\beta = -1.1 \pm 0.3$, similar to the expectation for a star formation or active galactic nuclei (AGN) origin \citep{2002ApJ...568...88Y, 1992ARA&A..30..575C}. However, the 10~GHz observations at $\sim 3623$ and $\sim 5480~$days post discovery demonstrate significant temporal evolution, with the radio source decreasing in brightness by a factor of $\sim 3$. This evolution is inconsistent with expectations for radio emission from star formation, though could be consistent with an AGN \citep{2011ApJ...742...49T, 2004AJ....127.2565D,2011ApJ...742...49T, 2011MNRAS.412..634B, 2016ApJ...818..105M}.} One way to test whether a galaxy harbors an AGN is to compare the colors of the galaxy in NASA's Wide-field Infrared Survey Explorer \citep[WISE, ][]{2010AJ....140.1868W}, but the host galaxy of PTF10tqv is too faint to be detected in WISE. While we cannot use WISE to rule out an AGN origin for the radio emission, the apparent offset of the radio source from the host galaxy center disfavors an AGN.

\added{We next asses the expected contribution to the radio emission from star formation. 
The star forming galaxies in the VLA-COSMOS source catalog at $z < 0.3$ have ${\rm SFR}_{\rm Radio}$ of $\lesssim 10 \times$ their optically derived ${\rm SFR}$ \citep{2017A&A...602A...2S}. Assuming ${\rm SFR}_{\rm Radio} \lesssim 10 \times$ that of the optically derived ${\rm SFR}$ of the host of PTF10tqv \citep[$0.022^{+0.066}_{-0.002}~M_{\odot}/{\rm yr}$,][]{2020ApJ...892..153M} we would expect the star formation contribution to the radio brightness at 3.1 GHz to be $\lesssim 12~\mu$Jy, $\gtrsim 8\times$ less than the measured value from the MeerKAT observation. } 
Given the offset, \added{temporal evolution}, and inconsistent radio derived ${\rm SFR}$, we consider it likely that the radio emission is related to the SN PTF10tqv, rather than the host galaxy. 

\subsubsection{Radio Emission from a Relativistic Jet}

Similar to Section~\ref{sec:Analysis_nondet}, we explore the allowed $n$-$E$ parameter space of PTF10tqv for the three $\theta_{\rm obs}$ considered, given our \added{3.1, 6, and 10 GHz detections}. \added{We utilize the radio observations to estimate $p$ to better tailor our model light curves to the data. At late times, the expected temporal and spectral behavior of radio emission from an off-axis jet is $F_{\nu} \propto t^{-p} \nu^{(1-p)/2}$. The 10~GHz emission of PTF10tqv follows $\alpha = -2.4 \pm 0.8$ and the 3.1--10~GHz emission at $\sim 5296$--$5480$ days follows $\beta = -1.1 \pm 0.3$. These values indicate $p = 2.4$--$3.2$, though we note this range is broad with large error bars. As a result we generate an additional grid of light curves with $p = 2.6$, leaving all other parameters the same (Section~\ref{sec:Models}). With these new models, we determine the $n$-$E$ parameter space consistent with the radio detections of PTF10tqv.} The result is displayed in Figure~\ref{fig:Energy_vs_Density_PTF10tqv}, where the models consistent within $2 \sigma$ of our detections are shaded for the three $\theta_{\rm obs}$\added{ (for simplicity, we omit the 6~GHz models in the light curve plot)}. \added{We find that our detections generally require
$10^{51} \lesssim E \lesssim 10^{51.7}~{\rm erg}$ in order to be consistent with our detections, with most of the solutions requiring $n \gtrsim 10^{-2.8}~{\rm cm}^{-3}$}. 
Additionally, \added{there exists a degeneracy at ${{n \gtrsim 10^{-0.5}~{\rm cm}^{-3}}}$, }
which is due to the lack of prior targeted radio observation at $\lesssim 10~$years. Finally, we note that there is no firm detected GRB counterpart to PTF10tqv\footnote{There was a short-duration (0.4\,s, $E_\mathrm{peak}=1\,$MeV) GRB detected by the \emph{Fermi} Gamma-ray Burst Monitor (GBM; \citealt{Meegan2009}) three days before the estimated time of first light \citep{2019A&A...621A..71T}, with the SN position at the 68\% containment region of the GBM localization. However, the position lies a degree outside the IPN-reported 3-$\sigma$ limits of 0.03 deg \citep{Palshin2013}, making them incompatible.} \citep{2019A&A...621A..71T}. Overall, the values of $E$ and $n$ that are consistent with our detections of PTF10tqv are similar to measured values from GRB afterglow studies \citep[e.g.][]{2002ApJ...571..779P, 2003ApJ...597..459Y, 2010ApJ...711..641C, 2011ApJ...732...29C, 2014ApJ...781....1L,  2015ApJ...799....3R, 2021ApJ...911...14K, 2022ApJ...940...53S}, indicating that the radio emission associated with PTF10tqv is consistent with a typical GRB jet viewed off-axis.

\subsubsection{Radio Emission from Other Mechanisms}

The PTF10tqv radio detection is also consistent with several other possible origins. The radio luminosity of PTF11qcj was an order of magnitude higher at a similar epoch and ultimately attributed to CSM interaction due to the inference of non-relativistic expansion speeds from VLBI observations \citep{2019ApJ...872..201P,2021ApJ...910...16P}. 
A pulsar wind nebula might produce radio emission with a similar luminosity and spectral index to PTF10tqv, as was argued for the radio transient VT 1137–0337 \citep{Dong2023}, \added{though the emission associated with PTF10tqv is fading at a significantly faster rate}. 
Finally, late-time radio emission from a mildly relativistic component (either associated with a GRB jet or SN ejecta) might be expected on these timescales \citep{2016MNRAS.461.1568K}. 
Ultimately, additional radio observations, including VLBI, are needed to distinguish between all of these scenarios.

\section{Discussion}
\label{sec:Discussion}

\begin{figure*}[!t]
\centering
\includegraphics[width=0.8\linewidth,clip=]{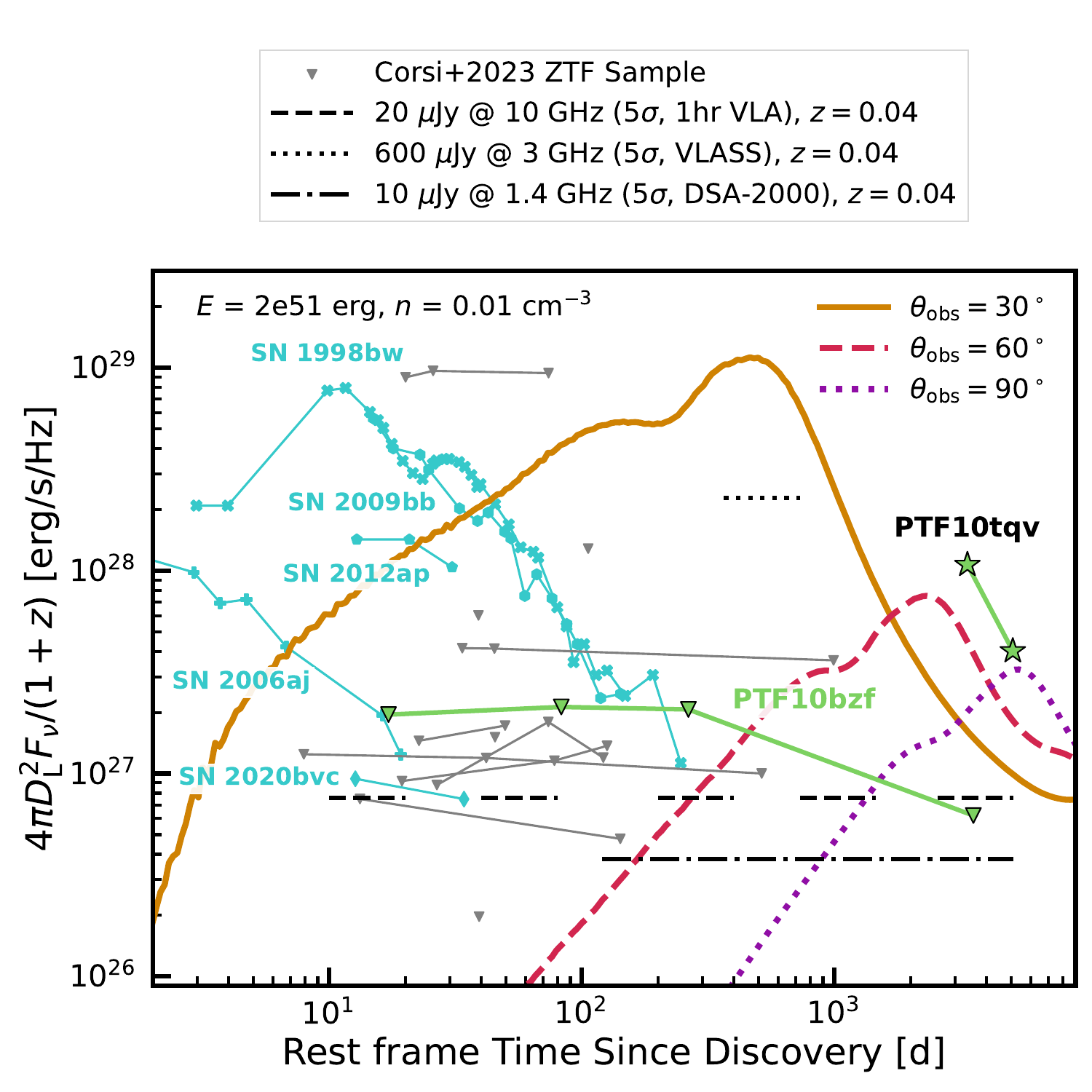} 
\caption{Rest frame light curves for 10~GHz afterglow models at off-axis angles of $\theta_{\rm obs} = 30, 60, 90^\circ$ (lines, orange, pink, purple, respectively) for $E = 2 \times 10^{51}~{\rm erg}$ and $n = 0.01~{\rm cm}^{-3}$ generated by the \texttt{FIREFLY} code \citep{2024ApJ...976..252D}. The light curves of PTF10bzf \citep[][this work, triangles]{2016ApJ...830...42C} and PTF10tqv (this work, star{s}) are shown in green.
Also shown are the radio non-detections presented in \citet{2023ApJ...953..179C} for a sample of ZTF SNe Ic-bl (grey, triangles represent $3\sigma$ upper limits), as well as the radio light curves of several supernova: SN\,1998bw \citep[associated with GRB\,980425,][]{1998Natur.395..663K, 1998ApJ...497..288W}, SN\,2006aj \citep[assoiciated with GRB\,060218, ][]{2006Natur.442.1014S}, SN\,2009bb \citep{2010Natur.463..513S}, SN\,2012ap \citep{2015ApJ...805..187C}, and SN\,2020bvc \citep{2020ApJ...902...86H}. The sampling of PTF10bzf represents a well-spaced observational strategy to constrain the existence of off-axis jets at any $\theta_{\rm obs}$. The $5 \sigma$ depth of a 1\,hr 10\,GHz VLA observation at our proposed cadence (black horizontal dashed lines), a 3~GHz VLASS observation at $\sim 1$--$2~{\rm year}$ (black horizontal dotted line) and a 1.4~GHz observation {with the Deep Synoptic Array 2000 (DSA-2000)} between $\sim 0.3$--$14~$year, at the median ZTF BTS Ic-bl redshift of $z = 0.04$ are presented for reference.}
\label{fig:ZTF_future}
\end{figure*}

Here, we explore the implications and limitations of our study, as well as paths forward to better detecting off-axis jets. Our candidate detection fraction is $1/14 \approx 7\%$ from this study. SNe Ic-bl are estimated to be about $\sim 3$--$5\%$ of the CC-SNe rate \citep{2020ApJ...904...35P, 2025arXiv250404393M}. Assuming a local volumetric rate of CC-SNe of $\sim 10^5~{\rm yr}^{-1}~{\rm Gpc}^{-3}$ \citep{2020ApJ...904...35P, 2025arXiv250404507M}, this implies a Ic-bl rate of $\sim{\rm few} \times 10^{3}~{\rm yr}^{-1}~{\rm Gpc}^{-3}$. In comparison, the local GRB rate is estimated to be $79^{+57}_{-33}~{\rm yr}^{-1}~{\rm Gpc}^{-3}$ \citep[beaming corrected,][]{2022ApJ...932...10G}. Given these rates for Ic-bl and GRBs, it is likely that $\sim 1$--$10\%$ of all Ic-bl produce successful GRB jets at any orientation \citep{2023ApJ...953..179C}. Therefore, our $\sim 7\%$ detection fraction is in line with the known rates.

However, converting this one candidate into a true rate of off-axis GRBs is not possible due to several limitations of our sample. In addition to the small sample size, the broad redshift range of the sample ($z = 0.015$--$0.158$) results in our observations having a broad range of constraints on the luminosity at late times ($\sim 10^{26}$--$10^{28}~{\rm erg~s}^{-1}~{\rm Hz}^{-1}$). This leads to weak constraints on the presence of lower energy ($\lesssim 10^{51}~{\rm erg}$) jets in lower density ($n \lesssim 10^{-1}~{\rm cm}^{-3}$) environments that are significantly off-axis ($\theta_{\rm obs} \gtrsim 60^{\circ}$). While a canonical long GRB is thought to have $E \approx 10^{51}~{\rm erg}$ \citep[e.g.][]{2001ApJ...562L..55F, 2003ApJ...594..674B, 2014PASA...31....8G, 2018IJMPD..2742002V}, the true $E$ range appears to vary over two orders of magnitude \citep[e.g.][]{2002ApJ...571..779P, 2003ApJ...597..459Y, 2010ApJ...711..641C, 2011ApJ...732...29C, 2014ApJ...781....1L,  2015ApJ...799....3R, 2021ApJ...911...14K, 2022ApJ...940...53S}, and therefore our observations may be insensitive to lower energy off-axis GRBs.  
Additionally, our assumptions about the microphysical parameters, in particular $\epsilon_{\rm B} = 0.1$, also may be optimistic compared to on-axis afterglow derived values \citep[e.g.][]{2002ApJ...571..779P, 2003ApJ...597..459Y, 2010ApJ...711..641C, 2011ApJ...732...29C, 2014ApJ...781....1L,  2015ApJ...799....3R, 2021ApJ...911...14K, 2022ApJ...940...53S}, and a factor of $10$ decrease in $\epsilon_{\rm B}$ results in a factor of $\sim 6$ decrease in luminosity. Therefore, if $\epsilon_{\rm B} \ll 0.1$, our constraints are weakened significantly. 
Our assumptions for $\theta_{\rm j}$, which we have fixed to $\sim 7^\circ$ in this work, also may affect our conclusions. In \citet{2020ApJ...892..153M}, the PTF Ic-bl sample that we explore in this paper was compared against as sample of GRB-SN.
Afterglow modeling of these GRB-SN have found a wide range of $\theta_{\rm j}$ \citep[e.g.][]{2006Natur.442.1014S, 2008A&A...480...35V, 2012ApJ...756..190Z, 2013ApJ...778...18M,2014A&A...567A..29M,2018ApJ...862...94L, 2021MNRAS.505.4106M, 2023MNRAS.525.5224M}, with several having $\theta_{\rm j} > 20^\circ$, much larger than our assumption. The choice of $\theta_{\rm j}\approx 7^\circ$ should not affect the \added{late-time} decline of our model light curves, but would affect the early time rise, and therefore should be explored further. \added{We have also assumed a constant density interstellar medium for our models, which while valid for our late-time observations when the jet radius is $\sim 10^{19}~$cm, may affect the conclusions of the earlier observations when the jet radius is $\lesssim 10^{17}~$cm. This effect should also be explored further.}
Finally, as mentioned previously, several of the SNe in our sample have only been observed once at $\sim 10~$years post discovery, allowing for faster fading jets to be feasible with our limited observations.

While many of the disadvantages stated above cannot be remedied retroactively, this study does lend insight into the future of detecting off-axis GRBs associated with SNe Ic-bl. For example, the ongoing ZTF Bright Transient Survey \citep[BTS, ][]{2020ApJ...895...32F} provides a well understood flux limited Ic-bl sample of $\sim 90$ events\footnote{\url{https://sites.astro.caltech.edu/ztf/bts/bts.php}}, with a median redshift of $\langle z \rangle = 0.04$. This large, nearby Ic-bl sample provides an ideal test set to search for off-axis jets.  Based on the GRB rates discussed above, we would expect $\sim 0.9$--$9$ of the ZTF BTS Ic-bl sample to harbor off-axis jets. Already, \citet{2023ApJ...953..179C} has performed such a test for 12 Ic-bl in this sample, with radio observations at $\sim 8$--$990~$days (rest frame), resulting in four radio counterpart detections, though no off-axis jets have been confirmed thus far. 

Given the lower redshift of this ZTF sample compared to the PTF sample, radio observations should be able to probe to luminosities on the order of $\sim 10^{27}~{\rm erg~s}^{-1}~{\rm Hz}^{-1}$ at 10~GHz with the VLA (Figure~\ref{fig:ZTF_future}). This will also allow for testing $\epsilon_{\rm B} \ll 0.1$ models without significantly weakening our constraints on the $n$-$E$ parameter space. To determine the best observational strategy, we use the PTF sample as a test case. As discussed, some of the events in the PTF sample were only observed once at $\sim 10~{\rm years}$ post discovery, resulting in weak constraints on the presence of a fast fading off-axis jet. In order to avoid a similar issue with the ZTF sample, well spaced observations of the ZTF sample should be conducted.
The best sampled Ic-bl in the PTF sample is PTF10bzf ($z = 0.0498$), with four epochs of radio observations spanning $\sim 17$--$3543~{\rm days}$ (rest frame, \citealt{2016ApJ...830...42C}, this work). Within our sample, this observational strategy results in some of the strongest constraints on the $n$-$E$ parameter space, and therefore we recommend following a similar strategy for the ZTF sample. We explore how this observational strategy will constrain the $n$-$E$ parameter space by assuming mock observations of a $z = 0.04$ Ic-bl with observations at rest frame times of $\sim 10, 40, 365, 1100, 3650~{\rm days}$ (similar to the observational strategy of PTF10bzf but including an observation at $\sim 3~{\rm years}$.) that reach depths of $\sim 7 \times 10^{26}~{\rm erg~s}^{-1}~{\rm Hz}^{-1}$. We find that with this observational strategy, we would be sensitive to off-axis jets of $\theta_{\rm obs} = 30^{\circ}$ 
($\theta_{\rm obs} \gtrsim 60^{\circ}$) with energies of $\gtrsim 10^{50}~{\rm erg}$ and ambient medium densities of $\gtrsim 10^{-3.1}~{\rm cm}^{-3}$ ($\gtrsim 10^{-1.4}~{\rm cm}^{-3}$)\added{, as demonstrated in Figure~\ref{fig:Energy_vs_Density}}. Additionally, this observational strategy would be able to constrain the presence of not only radio emission similar to GRB\,980424/SN\,1998bw-\citep{1998Natur.395..663K, 1998ApJ...497..288W}, but also the fast-fading GRB\,060218/SN\,2006aj \citep{2006ApJ...638..930S} and the  relativistic/mildly relativistic \added{SNe}\,2009bb, 2012ap, and 2020bvc \citep{2010Natur.463..513S, 2015ApJ...805..187C,2020ApJ...902...86H}.

While some ZTF BTS Ic-bl have had radio observations \citep{Ho2020,2020ApJ...902...86H,2023ApJ...953..179C,Srinivasaragavan2024} most ZTF BTS Ic-bl have not been observed, and those that have been observed are typically only monitored out to $\sim 100~{\rm days}$ (rest frame). However, in contrast to the PTF sample, VLASS observations of all ZTF objects can still place meaningful constraints on the presence of radio emission. The first epoch of VLASS was taken in 2017/2018 \citep{2020PASP..132c5001L}, roughly corresponding to the start of the ZTF sample, whereas this was $\sim 7$ years after some of the PTF events in our sample. Thanks to the shorter delay, VLASS images can be used to constrain early bright emission from off-axis jets of $\theta_{\rm obs} \lesssim 30^{\circ}$ for the ZTF sample (Figure~\ref{fig:ZTF_future}). Additionally, the 
\added{DSA-2000} will be coming online in the late 2020's \citep{2019BAAS...51g.255H}, imaging the radio sky at unprecedented depths (rms $\sim 2~\mu{\rm Jy}$ at 1.4~GHz) and cadences (every $\sim 120~$days). The DSA-2000 will not only probe radio emission from a ZTF Ic-bl at a deeper level than targeted VLA observations (Figure~\ref{fig:ZTF_future}), it will also be integral in characterizing radio emission from the host galaxies of these events.
These earlier observations from existing campaigns and VLASS will probe off-axis jets at $\theta_{\rm obs}\lesssim 60^\circ$, while targeted observations and observations from the DSA-2000 at $\sim 3$--$10~{\rm years}$ (rest frame) can probe jets at $\theta_{\rm obs} \approx 90^\circ$. Given these considerations, the ZTF sample provides the most promising opportunity for detecting off-axis jets from SNe Ic-bl in the near future, assuming methodical follow-up of the events is performed. Additionally, given the young age of the ZTF sample, VLBI observations of any events with high SNR detections should be performed in order to determine whether the radio emission is due to an off-axis jet or CSM interation, as was done for PTF11qcj  \citep{2019ApJ...872..201P} and FIRST J141918.9+394036 \citep{2022ApJ...924...16M}. 

\section{Summary and Conclusions}
\label{sec:Conclusion}

We presented late-time VLA radio observations of SNe Ic-bl that occurred in GRB-like host-galaxy environments, and that presumably had similar stripped massive star progenitors as long GRBs. We additionally used newly developed synchrotron radiation code to generate off-axis radio light curves, which we compared against our observations. Our observations probed significantly later times than past SNe Ic-bl radio campaigns, making them more sensitive to highly off-axis jets. 

We detected one object (PTF10tqv) with the VLA at 10~GHz ($\sim 3623~$years post discovery) resulting in a high luminosity of $\sim 10^{28}\,$erg\,s$^{-1}$\,Hz$^{-1}$. \added{Additional observations with MeerKAT at 3.1~GHz ($\sim 5296~$days post discovery) and the VLA at 6 and 10~GHz ($\sim 5480~$days post discovery) confirm the transient nature of the radio source.} Although multiple origins are possible, our detections of PTF10tqv are consistent with an off-axis jet with an energy of \added{${\sim 10^{51}}$--${10^{51.7}~}$erg}, similar to the inferred values for classical on-axis GRBs.  Additional radio observations will be needed to confirm the 
source size to determine whether this radio emission arises from an off-axis jet.

We find that the lack of radio detections in the other objects do not rule out off-axis jets entirely, but do imply that a ``typical'' long-duration GRB jet (with jet energy of $\approx 10^{51}~{\rm erg}$, and circumburst densities of $\gtrsim 10^{-1}~{\rm cm}^{-3}$) was not present in any of the other SNe in our sample, even for fully off-axis jets ($\theta_{\rm obs} = 90^\circ$). The reason for this may be that the jet in these events was choked in the stellar envelope, perhaps due to a shorter engine duration. Our detection fraction is consistent with relative rates of SNe Ic-bl and estimated collimation-corrected GRB rates in the local universe. 

Based on our work, we suggest a path forward for systematically discovering off-axis jets in the aftermath of SNe Ic-bl, taking advantage of the much larger samples of current optical surveys such as ZTF, and the deep radio observations that will be available with surveys such as the DSA-2000. These observations will be able to constrain the existence of off-axis GRBs with energies of $\gtrsim 10^{50}~{\rm erg}$ at circumburst densities of $\gtrsim 10^{-1.4}~{\rm cm}^{-3}$, depending on the microphysical parameters assumed. 

\begin{acknowledgements}

G.S. thanks Eric Burns and Adam Goldstein for their helpful discussions regarding {\it Fermi}.

G.S. and A.Y.Q.H. acknowledge support in part from a Sloan Research Fellowship (Award Number FG-2024-21320) from the Alfred P. Sloan Foundation. 
M.M. acknowledge support in part from ADAP program grant No. 80NSSC22K0486, from the NSF grant AST-2206657 and from the National Science Foundation under Cooperative Agreement 2421782 and the Simons Foundation grant MPS-AI-00010515 awarded to the NSF-Simons AI Institute for Cosmic Origins — CosmicAI, \url{https://www.cosmicai.org/}. A.C. acknowledges support from NASA/Swift Guest Investigator Programs and from the National Science Foundation (NSF/AST-2431072). Hydrodynamical calculations were carried out on the Petunia cluster at Purdue University. This project and the development of the code Firefly was supported by NASA under grant No. 80NSSC22K1615.

The National Radio Astronomy Observatory are facilities of the U.S. National Science Foundation operated under cooperative agreement by Associated Universities, Inc. The MeerKAT telescope is operated by the South African Radio Astronomy Observatory, which is a facility of the National Research Foundation, an agency of the Department of Science and Innovation.
LOFAR is the Low Frequency Array designed and constructed by ASTRON. It has observing, data processing, and data storage facilities in several countries, which are owned by various parties (each with their own funding sources), and which are collectively operated by the ILT foundation under a joint scientific policy. The ILT resources have benefited from the following recent major funding sources: CNRS-INSU, Observatoire de Paris and Université d'Orléans, France; BMBF, MIWF-NRW, MPG, Germany; Science Foundation Ireland (SFI), Department of Business, Enterprise and Innovation (DBEI), Ireland; NWO, The Netherlands; The Science and Technology Facilities Council, UK; Ministry of Science and Higher Education, Poland; The Istituto Nazionale di Astrofisica (INAF), Italy. This scientific work uses data obtained from Inyarrimanha Ilgari Bundara / the Murchison Radio-astronomy Observatory. We acknowledge the Wajarri Yamaji People as the Traditional Owners and native title holders of the Observatory site. CSIRO’s ASKAP radio telescope is part of the Australia Telescope National Facility (\url{https://ror.org/05qajvd42}). Operation of ASKAP is funded by the Australian Government with support from the National Collaborative Research Infrastructure Strategy. ASKAP uses the resources of the Pawsey Supercomputing Research Centre. Establishment of ASKAP, Inyarrimanha Ilgari Bundara, the CSIRO Murchison Radio-astronomy Observatory and the Pawsey Supercomputing Research Centre are initiatives of the Australian Government, with support from the Government of Western Australia and the Science and Industry Endowment Fund. This paper includes archived data obtained through the CSIRO ASKAP Science Data Archive, CASDA (\url{https://data.csiro.au}).

\end{acknowledgements}

\facilities{VLA, MeerKAT}
\software{CASA \citep{2007ASPC..376..127M, 2022PASP..134k4501C}, pwkit \citep{2017ascl.soft04001W}, matplotlib \citep{matplotlib}, pandas \citep{pandas}, numpy \citep{numpy}}

\appendix
\restartappendixnumbering
\renewcommand{\thefigure}{A.\arabic{figure}}
\setcounter{figure}{0} 

\begin{figure*}[!t]
\centering
\includegraphics[width=0.8\linewidth,clip=]{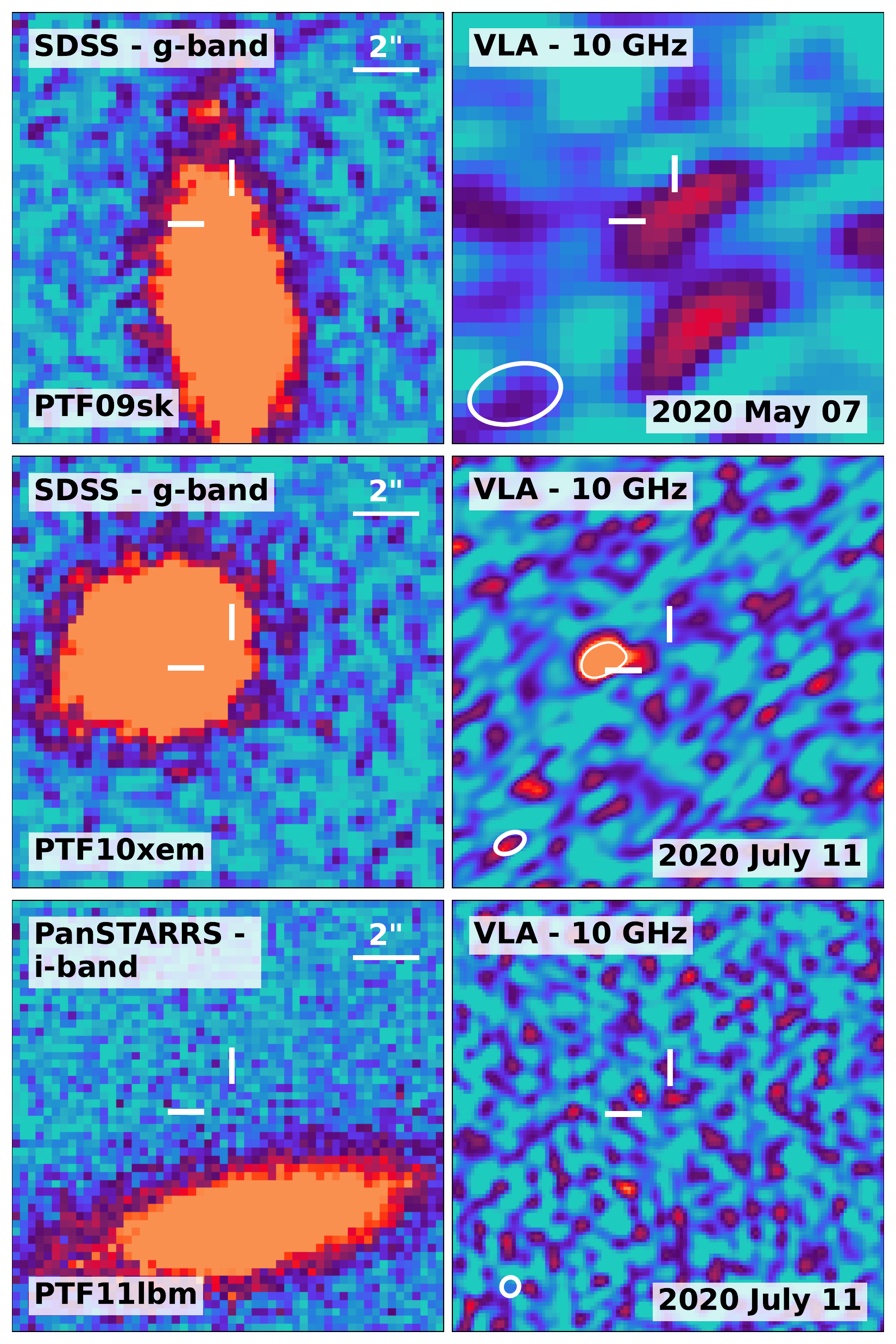} 
\caption{Host galaxy (left, SDSS $g$-band or PanSTARRS $i$-band) and VLA 10~GHz (right) images of PTF09sk, PTF10xem, PTF11lbm (top, middle, bottom, respectively). Crosshairs represent the optical location of each SN. For the VLA images, contours represent the $5\sigma$ level of the image. When available, the observatory and observing frequency/band is stated in the top left, the length scale is stated in the top right, the beam size is represented as an ellipse in the bottom left, and the observation date of is stated in the bottom right of each image. The radio detection in the PTF10xem VLA image is coincident with the center of its host galaxy, SHOC 084. PTF09sk and PTF11lbm both have $< 5 \sigma$ sources near the location of the SN, however association with the SN is not possible with available data (see Section~\ref{sec:Radio_Observations})}
\label{fig:SN_VLA_Images}
\end{figure*}

\begin{figure*}[!t]
\centering
\includegraphics[width=0.99\linewidth,clip=]{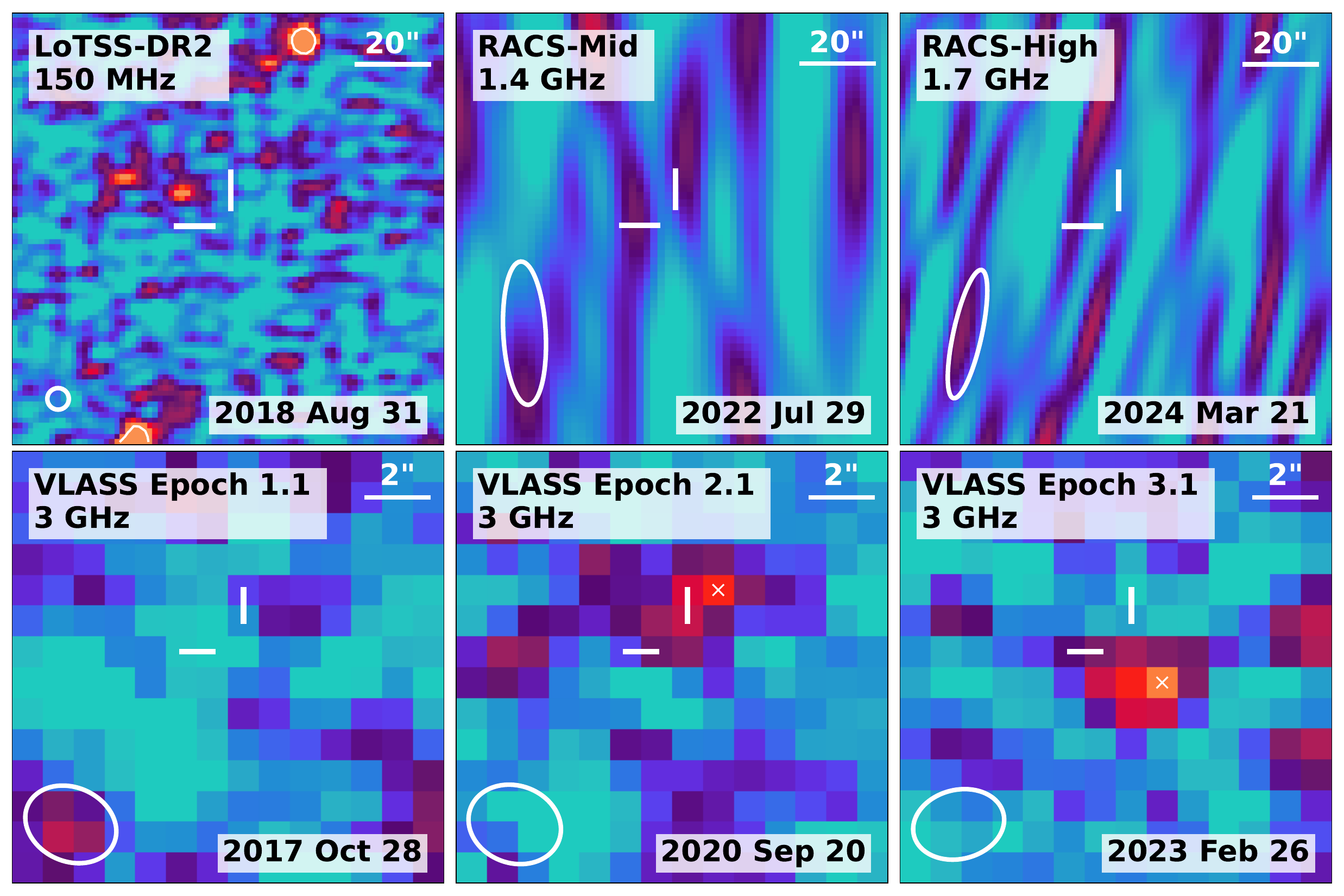} 
\caption{Radio survey images from LoTSS-DR2 (top left, \citealt{2022A&A...659A...1S}), RACS (top middle and right, \citealt{2023PASA...40...34D, 2024PASA...41....3D, 2025arXiv250104978D}), and VLASS (bottom, \citealt{2020PASP..132c5001L}) of the location of PTF11cmh. Crosshairs represent the optical location of the supernova. When applicable, contours represent the $5\sigma$ level of the image and crosses represent the location of the brightest pixel near the location of PTF11cmh (we note that the brightest pixel is below a $5\sigma$ detection). The observing frequency is stated in the top left, the length scale is stated in the top right, the beam size is represented as an ellipse in the bottom left, and the observation date is stated on the bottom right of each image.}
\label{fig:DetectionImage_PTF11cmh}
\end{figure*}

\begin{figure*}[!t]
\centering
\includegraphics[width=0.99\linewidth,clip=]{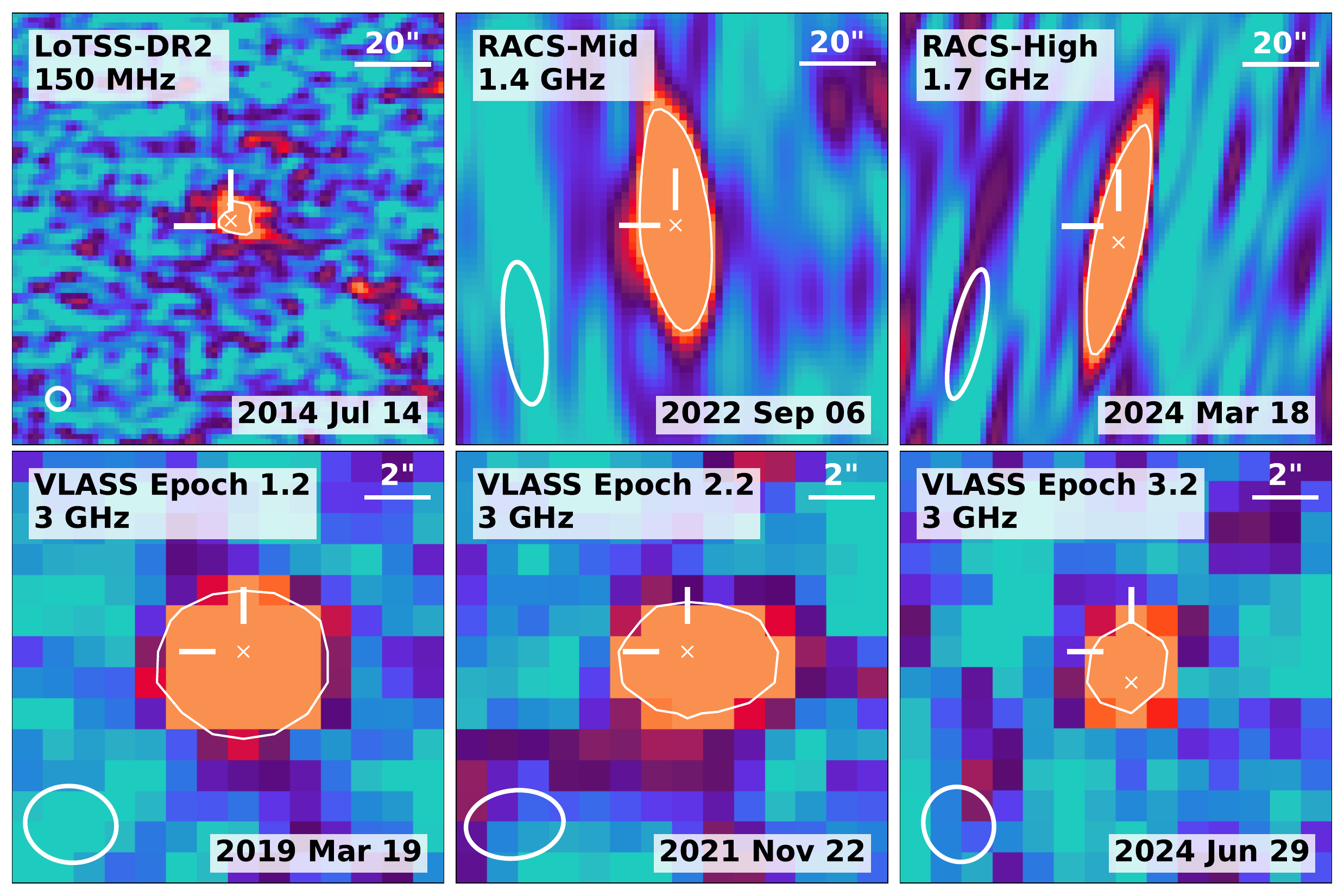} 
\caption{Same as Figure~\ref{fig:DetectionImage_PTF11cmh}, but for the location of PTF11qcj. The first VLASS detection (bottom left) was reported in \citet{2021ApJ...923L..24S}. The shape of the radio source in the LoTSS-DR2 image (top left) is inconsistent with the beam size of the observation, indicating at least some of the emission is not associated with PTF11cmh but rather its host galaxy (LEDA 2295826, \citealt{2021ApJ...908...75B}).}
\label{fig:DetectionImage_PTF11qcj}
\end{figure*}

In Figure~\ref{fig:SN_VLA_Images}, we show the optical host galaxy images and VLA radio images for PTF09sk, PTF10xem, and PTF11lbm. In Figure~\ref{fig:DetectionImage_PTF11cmh} we show the LoTTS-DR2, RACS-Mid, RAC-High, and VLASS images of PTF11cmh. We do the same for PTF11qcj in Figure~\ref{fig:DetectionImage_PTF11qcj}.

\newpage 
\bibliographystyle{aasjournalv7}
\bibliography{library,journals_apj}

\begin{thebibliography}{}
\expandafter\ifx\csname natexlab\endcsname\relax\def\natexlab#1{#1}\fi
\providecommand{\url}[1]{\href{#1}{#1}}
\providecommand{\dodoi}[1]{doi:~\href{http://doi.org/#1}{\nolinkurl{#1}}}
\providecommand{\doeprint}[1]{\href{http://ascl.net/#1}{\nolinkurl{http://ascl.net/#1}}}
\providecommand{\doarXiv}[1]{\href{https://arxiv.org/abs/#1}{\nolinkurl{https://arxiv.org/abs/#1}}}

\bibitem[{K.~W. {Bannister} {et~al.}(2011){Bannister}, {Murphy}, {Gaensler}, {Hunstead}, \& {Chatterjee}}]{2011MNRAS.412..634B}
{Bannister}, K.~W., {Murphy}, T., {Gaensler}, B.~M., {Hunstead}, R.~W., \& {Chatterjee}, S. 2011, \bibinfo{title}{{A 22-yr southern sky survey for transient and variable radio sources using the Molonglo Observatory Synthesis Telescope},} \mnras, 412, 634, \dodoi{10.1111/j.1365-2966.2010.17938.x}

\bibitem[{J. {Barnes} {et~al.}(2018){Barnes}, {Duffell}, {Liu}, {Modjaz}, {Bianco}, {Kasen}, \& {MacFadyen}}]{2018ApJ...860...38B}
{Barnes}, J., {Duffell}, P.~C., {Liu}, Y., {et~al.} 2018, \bibinfo{title}{{A GRB and Broad-lined Type Ic Supernova from a Single Central Engine},} \apj, 860, 38, \dodoi{10.3847/1538-4357/aabf84}

\bibitem[{R. {Barniol Duran} \& D. {Giannios}(2015){Barniol Duran} \& {Giannios}}]{2015MNRAS.454.1711B}
{Barniol Duran}, R., \& {Giannios}, D. 2015, \bibinfo{title}{{Radio rebrightening of the GRB afterglow by the accompanying supernova},} \mnras, 454, 1711, \dodoi{10.1093/mnras/stv2004}

\bibitem[{E. Berger {et~al.}(2003)Berger, Kulkarni, Frail, \& Soderberg}]{2003ApJ...599..408B}
Berger, E., Kulkarni, S.~R., Frail, D.~A., \& Soderberg, A.~M. 2003, \bibinfo{title}{A {Radio} {Survey} of {Type} {Ib} and {Ic} {Supernovae}: {Searching} for {Engine}-driven {Supernovae},} \apj, 599, 408, \dodoi{10.1086/379214}

\bibitem[{M.~F. {Bietenholz} {et~al.}(2021){Bietenholz}, {Bartel}, {Argo}, {Dua}, {Ryder}, \& {Soderberg}}]{2021ApJ...908...75B}
{Bietenholz}, M.~F., {Bartel}, N., {Argo}, M., {et~al.} 2021, \bibinfo{title}{{The Radio Luminosity-risetime Function of Core-collapse Supernovae},} \apj, 908, 75, \dodoi{10.3847/1538-4357/abccd9}

\bibitem[{M.~F. Bietenholz {et~al.}(2014)Bietenholz, De~Colle, Granot, Bartel, \& Soderberg}]{2014MNRAS.440..821B}
Bietenholz, M.~F., De~Colle, F., Granot, J., Bartel, N., \& Soderberg, A.~M. 2014, \bibinfo{title}{Radio limits on off-axis {GRB} afterglows and {VLBI} observations of {SN} 2003gk,} \mnras, 440, 821, \dodoi{10.1093/mnras/stu246}

\bibitem[{J.~S. {Bloom} {et~al.}(2003){Bloom}, {Frail}, \& {Kulkarni}}]{2003ApJ...594..674B}
{Bloom}, J.~S., {Frail}, D.~A., \& {Kulkarni}, S.~R. 2003, \bibinfo{title}{{Gamma-Ray Burst Energetics and the Gamma-Ray Burst Hubble Diagram: Promises and Limitations},} \apj, 594, 674, \dodoi{10.1086/377125}

\bibitem[{J.~N. {Bregman} {et~al.}(1986){Bregman}, {Glassgold}, {Huggins}, {Neugebauer}, {Soifer}, {Matthews}, {Elias}, {Webb}, {Pollock}, {Pica}, {Leacock}, {Smith}, {Aller}, {Aller}, {Hodge}, {Dent}, {Balonek}, {Barvainis}, {Roellig}, {Wisniewski}, {Rieke}, {Lebofsky}, {Wills}, {Wills}, {Ku}, {Bregman}, {Witteborn}, {Lester}, {Impey}, \& {Hackwell}}]{1986ApJ...301..708B}
{Bregman}, J.~N., {Glassgold}, A.~E., {Huggins}, P.~J., {et~al.} 1986, \bibinfo{title}{{Multifrequency Observations of the Superluminal Quasar 3C 345},} \apj, 301, 708, \dodoi{10.1086/163938}

\bibitem[{O. {Bromberg} {et~al.}(2012){Bromberg}, {Nakar}, {Piran}, \& {Sari}}]{2012ApJ...749..110B}
{Bromberg}, O., {Nakar}, E., {Piran}, T., \& {Sari}, R. 2012, \bibinfo{title}{{An Observational Imprint of the Collapsar Model of Long Gamma-Ray Bursts},} \apj, 749, 110, \dodoi{10.1088/0004-637X/749/2/110}

\bibitem[{S. {Campana} {et~al.}(2006){Campana}, {Mangano}, {Blustin}, {Brown}, {Burrows}, {Chincarini}, {Cummings}, {Cusumano}, {Della Valle}, {Malesani}, {M{\'e}sz{\'a}ros}, {Nousek}, {Page}, {Sakamoto}, {Waxman}, {Zhang}, {Dai}, {Gehrels}, {Immler}, {Marshall}, {Mason}, {Moretti}, {O'Brien}, {Osborne}, {Page}, {Romano}, {Roming}, {Tagliaferri}, {Cominsky}, {Giommi}, {Godet}, {Kennea}, {Krimm}, {Angelini}, {Barthelmy}, {Boyd}, {Palmer}, {Wells}, \& {White}}]{2006Natur.442.1008C}
{Campana}, S., {Mangano}, V., {Blustin}, A.~J., {et~al.} 2006, \bibinfo{title}{{The association of GRB 060218 with a supernova and the evolution of the shock wave},} \nat, 442, 1008, \dodoi{10.1038/nature04892}

\bibitem[{Z. {Cano} {et~al.}(2017){Cano}, {Wang}, {Dai}, \& {Wu}}]{Cano2017}
{Cano}, Z., {Wang}, S.-Q., {Dai}, Z.-G., \& {Wu}, X.-F. 2017, \bibinfo{title}{{The Observer's Guide to the Gamma-Ray Burst Supernova Connection},} Advances in Astronomy, 2017, 8929054, \dodoi{10.1155/2017/8929054}

\bibitem[{ {CASA Team} {et~al.}(2022){CASA Team}, {Bean}, {Bhatnagar}, {Castro}, {Donovan Meyer}, {Emonts}, {Garcia}, {Garwood}, {Golap}, {Gonzalez Villalba}, {Harris}, {Hayashi}, {Hoskins}, {Hsieh}, {Jagannathan}, {Kawasaki}, {Keimpema}, {Kettenis}, {Lopez}, {Marvil}, {Masters}, {McNichols}, {Mehringer}, {Miel}, {Moellenbrock}, {Montesino}, {Nakazato}, {Ott}, {Petry}, {Pokorny}, {Raba}, {Rau}, {Schiebel}, {Schweighart}, {Sekhar}, {Shimada}, {Small}, {Steeb}, {Sugimoto}, {Suoranta}, {Tsutsumi}, {van Bemmel}, {Verkouter}, {Wells}, {Xiong}, {Szomoru}, {Griffith}, {Glendenning}, \& {Kern}}]{2022PASP..134k4501C}
{CASA Team}, {Bean}, B., {Bhatnagar}, S., {et~al.} 2022, \bibinfo{title}{{CASA, the Common Astronomy Software Applications for Radio Astronomy},} \pasp, 134, 114501, \dodoi{10.1088/1538-3873/ac9642}

\bibitem[{S.~B. {Cenko} {et~al.}(2010){Cenko}, {Frail}, {Harrison}, {Kulkarni}, {Nakar}, {Chandra}, {Butler}, {Fox}, {Gal-Yam}, {Kasliwal}, {Kelemen}, {Moon}, {Ofek}, {Price}, {Rau}, {Soderberg}, {Teplitz}, {Werner}, {Bock}, {Bloom}, {Starr}, {Filippenko}, {Chevalier}, {Gehrels}, {Nousek}, \& {Piran}}]{2010ApJ...711..641C}
{Cenko}, S.~B., {Frail}, D.~A., {Harrison}, F.~A., {et~al.} 2010, \bibinfo{title}{{The Collimation and Energetics of the Brightest Swift Gamma-ray Bursts},} \apj, 711, 641, \dodoi{10.1088/0004-637X/711/2/641}

\bibitem[{S.~B. {Cenko} {et~al.}(2011){Cenko}, {Frail}, {Harrison}, {Haislip}, {Reichart}, {Butler}, {Cobb}, {Cucchiara}, {Berger}, {Bloom}, {Chandra}, {Fox}, {Perley}, {Prochaska}, {Filippenko}, {Glazebrook}, {Ivarsen}, {Kasliwal}, {Kulkarni}, {LaCluyze}, {Lopez}, {Morgan}, {Pettini}, \& {Rana}}]{2011ApJ...732...29C}
{Cenko}, S.~B., {Frail}, D.~A., {Harrison}, F.~A., {et~al.} 2011, \bibinfo{title}{{Afterglow Observations of Fermi Large Area Telescope Gamma-ray Bursts and the Emerging Class of Hyper-energetic Events},} \apj, 732, 29, \dodoi{10.1088/0004-637X/732/1/29}

\bibitem[{S. {Chakraborti} {et~al.}(2015){Chakraborti}, {Soderberg}, {Chomiuk}, {Kamble}, {Yadav}, {Ray}, {Hurley}, {Margutti}, {Milisavljevic}, {Bietenholz}, {Brunthaler}, {Pignata}, {Pian}, {Mazzali}, {Fransson}, {Bartel}, {Hamuy}, {Levesque}, {MacFadyen}, {Dittmann}, {Krauss}, {Briggs}, {Connaughton}, {Yamaoka}, {Takahashi}, {Ohno}, {Fukazawa}, {Tashiro}, {Terada}, {Murakami}, {Goldsten}, {Barthelmy}, {Gehrels}, {Cummings}, {Krimm}, {Palmer}, {Golenetskii}, {Aptekar}, {Frederiks}, {Svinkin}, {Cline}, {Mitrofanov}, {Golovin}, {Litvak}, {Sanin}, {Boynton}, {Fellows}, {Harshman}, {Enos}, {von Kienlin}, {Rau}, {Zhang}, \& {Savchenko}}]{2015ApJ...805..187C}
{Chakraborti}, S., {Soderberg}, A., {Chomiuk}, L., {et~al.} 2015, \bibinfo{title}{{A Missing-link in the Supernova-GRB Connection: The Case of SN 2012ap},} \apj, 805, 187, \dodoi{10.1088/0004-637X/805/2/187}

\bibitem[{J.~J. {Condon}(1992){Condon}}]{1992ARA&A..30..575C}
{Condon}, J.~J. 1992, \bibinfo{title}{{Radio emission from normal galaxies.},} \araa, 30, 575, \dodoi{10.1146/annurev.aa.30.090192.003043}

\bibitem[{A. {Corsi} {et~al.}(2012){Corsi}, {Ofek}, {Gal-Yam}, {Frail}, {Poznanski}, {Mazzali}, {Kulkarni}, {Kasliwal}, {Arcavi}, {Ben-Ami}, {Cenko}, {Filippenko}, {Fox}, {Horesh}, {Howell}, {Kleiser}, {Nakar}, {Rabinak}, {Sari}, {Silverman}, {Xu}, {Bloom}, {Law}, {Nugent}, \& {Quimby}}]{2012ApJ...747L...5C}
{Corsi}, A., {Ofek}, E.~O., {Gal-Yam}, A., {et~al.} 2012, \bibinfo{title}{{Evidence for a Compact Wolf-Rayet Progenitor for the Type Ic Supernova PTF 10vgv},} \apjl, 747, L5, \dodoi{10.1088/2041-8205/747/1/L5}

\bibitem[{A. {Corsi} {et~al.}(2014){Corsi}, {Ofek}, {Gal-Yam}, {Frail}, {Kulkarni}, {Fox}, {Kasliwal}, {Sullivan}, {Horesh}, {Carpenter}, {Maguire}, {Arcavi}, {Cenko}, {Cao}, {Mooley}, {Pan}, {Sesar}, {Sternberg}, {Xu}, {Bersier}, {James}, {Bloom}, \& {Nugent}}]{2014ApJ...782...42C}
{Corsi}, A., {Ofek}, E.~O., {Gal-Yam}, A., {et~al.} 2014, \bibinfo{title}{{A Multi-wavelength Investigation of the Radio-loud Supernova PTF11qcj and its Circumstellar Environment},} \apj, 782, 42, \dodoi{10.1088/0004-637X/782/1/42}

\bibitem[{A. {Corsi} {et~al.}(2016){Corsi}, {Gal-Yam}, {Kulkarni}, {Frail}, {Mazzali}, {Cenko}, {Kasliwal}, {Cao}, {Horesh}, {Palliyaguru}, {Perley}, {Laher}, {Taddia}, {Leloudas}, {Maguire}, {Nugent}, {Sollerman}, \& {Sullivan}}]{2016ApJ...830...42C}
{Corsi}, A., {Gal-Yam}, A., {Kulkarni}, S.~R., {et~al.} 2016, \bibinfo{title}{{Radio Observations of a Sample of Broad-line Type IC Supernovae Discovered by PTF/IPTF: A Search for Relativistic Explosions},} \apj, 830, 42, \dodoi{10.3847/0004-637X/830/1/42}

\bibitem[{A. Corsi {et~al.}(2023)Corsi, Ho, Cenko, Kulkarni, Anand, Yang, Sollerman, Srinivasaragavan, Omand, Balasubramanian, Frail, Fremling, Perley, Yao, Dahiwale, De, Dugas, Hankins, Jencson, Kasliwal, Tzanidakis, Bellm, Laher, Masci, Purdum, \& Regnault}]{2023ApJ...953..179C}
Corsi, A., Ho, A. Y.~Q., Cenko, S.~B., {et~al.} 2023, \bibinfo{title}{A {Search} for {Relativistic} {Ejecta} in a {Sample} of {ZTF} {Broad}-lined {Type} {Ic} {Supernovae},} \apj, 953, 179, \dodoi{10.3847/1538-4357/acd3f2}

\bibitem[{R.~G. {Dastidar} \& P.~C. {Duffell}(2024){Dastidar} \& {Duffell}}]{2024ApJ...976..252D}
{Dastidar}, R.~G., \& {Duffell}, P.~C. 2024, \bibinfo{title}{{Could the Recent Rebrightening of the GW170817A Afterglow Be Caused by a Counterjet?},} \apj, 976, 252, \dodoi{10.3847/1538-4357/ad86bf}

\bibitem[{W.~H. {de Vries} {et~al.}(2004){de Vries}, {Becker}, {White}, \& {Helfand}}]{2004AJ....127.2565D}
{de Vries}, W.~H., {Becker}, R.~H., {White}, R.~L., \& {Helfand}, D.~J. 2004, \bibinfo{title}{{Optical Properties of faint FIRST Variable Radio Sources},} \aj, 127, 2565, \dodoi{10.1086/383550}

\bibitem[{D.~Z. {Dong} \& G. {Hallinan}(2023){Dong} \& {Hallinan}}]{Dong2023}
{Dong}, D.~Z., \& {Hallinan}, G. 2023, \bibinfo{title}{{A Flat-spectrum Radio Transient at 122 Mpc Consistent with an Emerging Pulsar Wind Nebula},} \apj, 948, 119, \dodoi{10.3847/1538-4357/acc06c}

\bibitem[{S.~W. {Duchesne} {et~al.}(2023){Duchesne}, {Thomson}, {Pritchard}, {Lenc}, {Moss}, {McConnell}, {Wieringa}, {Whiting}, {Wang}, {Wang}, {Rose}, {Raja}, {Murphy}, {Leung}, {Huynh}, {Hotan}, {Hodgson}, \& {Heald}}]{2023PASA...40...34D}
{Duchesne}, S.~W., {Thomson}, A.~J.~M., {Pritchard}, J., {et~al.} 2023, \bibinfo{title}{{The Rapid ASKAP Continuum Survey IV: continuum imaging at 1367.5 MHz and the first data release of RACS-mid},} \pasa, 40, e034, \dodoi{10.1017/pasa.2023.31}

\bibitem[{S.~W. {Duchesne} {et~al.}(2024){Duchesne}, {Grundy}, {Heald}, {Lenc}, {Leung}, {McConnell}, {Murphy}, {Pritchard}, {Rose}, {Thomson}, {Wang}, {Wang}, \& {Whiting}}]{2024PASA...41....3D}
{Duchesne}, S.~W., {Grundy}, J.~A., {Heald}, G.~H., {et~al.} 2024, \bibinfo{title}{{The Rapid ASKAP Continuum Survey V: Cataloguing the sky at 1 367.5 MHz and the second data release of RACS-mid},} \pasa, 41, e003, \dodoi{10.1017/pasa.2023.60}

\bibitem[{S.~W. {Duchesne} {et~al.}(2025){Duchesne}, {Ross}, {Thomson}, {Lenc}, {Murphy}, {Galvin}, {Hotan}, {Moss}, \& {Whiting}}]{2025arXiv250104978D}
{Duchesne}, S.~W., {Ross}, K., {Thomson}, A.~J.~M., {et~al.} 2025, \bibinfo{title}{{The Rapid ASKAP Continuum Survey (RACS) VI: The RACS-high 1655.5 MHz images and catalogue},} arXiv e-prints, arXiv:2501.04978, \dodoi{10.48550/arXiv.2501.04978}

\bibitem[{P.~C. Duffell \& A.~I. MacFadyen(2013)Duffell \& MacFadyen}]{Duffell+2013}
Duffell, P.~C., \& MacFadyen, A.~I. 2013, \bibinfo{title}{A “BOOSTED FIREBALL” MODEL FOR STRUCTURED RELATIVISTIC JETS,} \apjl, 776, L9, \dodoi{10.1088/2041-8205/776/1/L9}

\bibitem[{P.~C. Duffell \& A.~I. MacFadyen(2015)Duffell \& MacFadyen}]{Duffell+2015}
Duffell, P.~C., \& MacFadyen, A.~I. 2015, \bibinfo{title}{FROM ENGINE TO AFTERGLOW: COLLAPSARS NATURALLY PRODUCE TOP-HEAVY JETS AND EARLY-TIME PLATEAUS IN GAMMA-RAY BURST AFTERGLOWS,} \apj, 806, 205, \dodoi{10.1088/0004-637X/806/2/205}

\bibitem[{D.~A. {Frail} {et~al.}(2001){Frail}, {Kulkarni}, {Sari}, {Djorgovski}, {Bloom}, {Galama}, {Reichart}, {Berger}, {Harrison}, {Price}, {Yost}, {Diercks}, {Goodrich}, \& {Chaffee}}]{2001ApJ...562L..55F}
{Frail}, D.~A., {Kulkarni}, S.~R., {Sari}, R., {et~al.} 2001, \bibinfo{title}{{Beaming in Gamma-Ray Bursts: Evidence for a Standard Energy Reservoir},} \apjl, 562, L55, \dodoi{10.1086/338119}

\bibitem[{C. {Fremling} {et~al.}(2020){Fremling}, {Miller}, {Sharma}, {Dugas}, {Perley}, {Taggart}, {Sollerman}, {Goobar}, {Graham}, {Neill}, {Nordin}, {Rigault}, {Walters}, {Andreoni}, {Bagdasaryan}, {Belicki}, {Cannella}, {Bellm}, {Cenko}, {De}, {Dekany}, {Frederick}, {Golkhou}, {Graham}, {Helou}, {Ho}, {Kasliwal}, {Kupfer}, {Laher}, {Mahabal}, {Masci}, {Riddle}, {Rusholme}, {Schulze}, {Shupe}, {Smith}, {van Velzen}, {Yan}, {Yao}, {Zhuang}, \& {Kulkarni}}]{2020ApJ...895...32F}
{Fremling}, C., {Miller}, A.~A., {Sharma}, Y., {et~al.} 2020, \bibinfo{title}{{The Zwicky Transient Facility Bright Transient Survey. I. Spectroscopic Classification and the Redshift Completeness of Local Galaxy Catalogs},} \apj, 895, 32, \dodoi{10.3847/1538-4357/ab8943}

\bibitem[{A.~S. {Fruchter} {et~al.}(2006){Fruchter}, {Levan}, {Strolger}, {Vreeswijk}, {Thorsett}, {Bersier}, {Burud}, {Castro Cer{\'o}n}, {Castro-Tirado}, {Conselice}, {Dahlen}, {Ferguson}, {Fynbo}, {Garnavich}, {Gibbons}, {Gorosabel}, {Gull}, {Hjorth}, {Holland}, {Kouveliotou}, {Levay}, {Livio}, {Metzger}, {Nugent}, {Petro}, {Pian}, {Rhoads}, {Riess}, {Sahu}, {Smette}, {Tanvir}, {Wijers}, \& {Woosley}}]{Fruchter2006}
{Fruchter}, A.~S., {Levan}, A.~J., {Strolger}, L., {et~al.} 2006, \bibinfo{title}{{Long {\ensuremath{\gamma}}-ray bursts and core-collapse supernovae have different environments},} \nat, 441, 463, \dodoi{10.1038/nature04787}

\bibitem[{G. {Ghirlanda} \& R. {Salvaterra}(2022){Ghirlanda} \& {Salvaterra}}]{2022ApJ...932...10G}
{Ghirlanda}, G., \& {Salvaterra}, R. 2022, \bibinfo{title}{{The Cosmic History of Long Gamma-Ray Bursts},} \apj, 932, 10, \dodoi{10.3847/1538-4357/ac6e43}

\bibitem[{J. {Granot} {et~al.}(2002){Granot}, {Panaitescu}, {Kumar}, \& {Woosley}}]{Granot2002}
{Granot}, J., {Panaitescu}, A., {Kumar}, P., \& {Woosley}, S.~E. 2002, \bibinfo{title}{{Off-Axis Afterglow Emission from Jetted Gamma-Ray Bursts},} \apjl, 570, L61, \dodoi{10.1086/340991}

\bibitem[{J. {Granot} \& A.~J. {van der Horst}(2014){Granot} \& {van der Horst}}]{2014PASA...31....8G}
{Granot}, J., \& {van der Horst}, A.~J. 2014, \bibinfo{title}{{Gamma-Ray Burst Jets and their Radio Observations},} \pasa, 31, e008, \dodoi{10.1017/pasa.2013.44}

\bibitem[{J. {Greiner} {et~al.}(2016){Greiner}, {Micha{\l}owski}, {Klose}, {Hunt}, {Gentile}, {Kamphuis}, {Herrero-Illana}, {Wieringa}, {Kr{\"u}hler}, {Schady}, {Elliott}, {Graham}, {Ibar}, {Knust}, {Nicuesa Guelbenzu}, {Palazzi}, {Rossi}, \& {Savaglio}}]{2016A&A...593A..17G}
{Greiner}, J., {Micha{\l}owski}, M.~J., {Klose}, S., {et~al.} 2016, \bibinfo{title}{{Probing dust-obscured star formation in the most massive gamma-ray burst host galaxies},} \aap, 593, A17, \dodoi{10.1051/0004-6361/201628861}

\bibitem[{G. {Hallinan} {et~al.}(2019){Hallinan}, {Ravi}, {Weinreb}, {Kocz}, {Huang}, {Woody}, {Lamb}, {D'Addario}, {Catha}, {Law}, {Kulkarni}, {Phinney}, {Eastwood}, {Bouman}, {McLaughlin}, {Ransom}, {Siemens}, {Cordes}, {Lynch}, {Kaplan}, {Brazier}, {Bhatnagar}, {Myers}, {Walter}, \& {Gaensler}}]{2019BAAS...51g.255H}
{Hallinan}, G., {Ravi}, V., {Weinreb}, S., {et~al.} 2019, \bibinfo{title}{{The DSA-2000 {\textemdash} A Radio Survey Camera},} in Bulletin of the American Astronomical Society, Vol.~51, 255, \dodoi{10.48550/arXiv.1907.07648}

\bibitem[{A.~Y.~Q. {Ho} {et~al.}(2020{\natexlab{a}}){Ho}, {Corsi}, {Cenko}, {Taddia}, {Kulkarni}, {Adams}, {De}, {Dekany}, {Frederiks}, {Fremling}, {Golkhou}, {Graham}, {Hung}, {Kupfer}, {Laher}, {Mahabal}, {Masci}, {Miller}, {Neill}, {Reiley}, {Riddle}, {Ridnaia}, {Rusholme}, {Sharma}, {Sollerman}, {Soumagnac}, {Svinkin}, \& {Shupe}}]{Ho2020}
{Ho}, A. Y.~Q., {Corsi}, A., {Cenko}, S.~B., {et~al.} 2020{\natexlab{a}}, \bibinfo{title}{{The Broad-lined Ic Supernova ZTF18aaqjovh (SN 2018bvw): An Optically Discovered Engine-driven Supernova Candidate with Luminous Radio Emission},} \apj, 893, 132, \dodoi{10.3847/1538-4357/ab7f3b}

\bibitem[{A.~Y.~Q. {Ho} {et~al.}(2020{\natexlab{b}}){Ho}, {Kulkarni}, {Perley}, {Cenko}, {Corsi}, {Schulze}, {Lunnan}, {Sollerman}, {Gal-Yam}, {Anand}, {Barbarino}, {Bellm}, {Bruch}, {Burns}, {De}, {Dekany}, {Delacroix}, {Duev}, {Frederiks}, {Fremling}, {Goldstein}, {Golkhou}, {Graham}, {Hale}, {Kasliwal}, {Kupfer}, {Laher}, {Martikainen}, {Masci}, {Neill}, {Ridnaia}, {Rusholme}, {Savchenko}, {Shupe}, {Soumagnac}, {Strotjohann}, {Svinkin}, {Taggart}, {Tartaglia}, {Yan}, \& {Zolkower}}]{2020ApJ...902...86H}
{Ho}, A. Y.~Q., {Kulkarni}, S.~R., {Perley}, D.~A., {et~al.} 2020{\natexlab{b}}, \bibinfo{title}{{SN 2020bvc: A Broad-line Type Ic Supernova with a Double-peaked Optical Light Curve and a Luminous X-Ray and Radio Counterpart},} \apj, 902, 86, \dodoi{10.3847/1538-4357/aba630}

\bibitem[{J.~D. Hunter(2007)Hunter}]{matplotlib}
Hunter, J.~D. 2007, \bibinfo{title}{Matplotlib: A 2D Graphics Environment,} Computing in Science Engineering, 9, 90, \dodoi{10.1109/MCSE.2007.55}

\bibitem[{C.~M. {Irwin} \& R.~A. {Chevalier}(2016){Irwin} \& {Chevalier}}]{Irwin2016}
{Irwin}, C.~M., \& {Chevalier}, R.~A. 2016, \bibinfo{title}{{Jet or shock breakout? The low-luminosity GRB 060218},} \mnras, 460, 1680, \dodoi{10.1093/mnras/stw1058}

\bibitem[{C.~M. {Irwin} \& K. {Hotokezaka}(2024){Irwin} \& {Hotokezaka}}]{2024arXiv241206736I}
{Irwin}, C.~M., \& {Hotokezaka}, K. 2024, \bibinfo{title}{{Revisiting GRB 060218: new insights into low-luminosity gamma-ray bursts from a revised shock breakout model},} arXiv e-prints, arXiv:2412.06736, \dodoi{10.48550/arXiv.2412.06736}

\bibitem[{L. {Izzo} {et~al.}(2020){Izzo}, {Auchettl}, {Hjorth}, {De Colle}, {Gall}, {Angus}, {Raimundo}, \& {Ramirez-Ruiz}}]{Izzo2020}
{Izzo}, L., {Auchettl}, K., {Hjorth}, J., {et~al.} 2020, \bibinfo{title}{{Broad-line type Ic SN 2020bvc. Signatures of an off-axis gamma-ray burst afterglow},} \aap, 639, L11, \dodoi{10.1051/0004-6361/202038152}

\bibitem[{T. {Kangas} \& A.~S. {Fruchter}(2021){Kangas} \& {Fruchter}}]{2021ApJ...911...14K}
{Kangas}, T., \& {Fruchter}, A.~S. 2021, \bibinfo{title}{{The Late-time Radio Behavior of Gamma-ray Burst Afterglows: Testing the Standard Model},} \apj, 911, 14, \dodoi{10.3847/1538-4357/abe76b}

\bibitem[{A. {Kathirgamaraju} {et~al.}(2016){Kathirgamaraju}, {Barniol Duran}, \& {Giannios}}]{2016MNRAS.461.1568K}
{Kathirgamaraju}, A., {Barniol Duran}, R., \& {Giannios}, D. 2016, \bibinfo{title}{{GRB off-axis afterglows and the emission from the accompanying supernovae},} \mnras, 461, 1568, \dodoi{10.1093/mnras/stw1441}

\bibitem[{S.~R. {Kulkarni} {et~al.}(1998){Kulkarni}, {Frail}, {Wieringa}, {Ekers}, {Sadler}, {Wark}, {Higdon}, {Phinney}, \& {Bloom}}]{1998Natur.395..663K}
{Kulkarni}, S.~R., {Frail}, D.~A., {Wieringa}, M.~H., {et~al.} 1998, \bibinfo{title}{{Radio emission from the unusual supernova 1998bw and its association with the {\ensuremath{\gamma}}-ray burst of 25 April 1998},} \nat, 395, 663, \dodoi{10.1038/27139}

\bibitem[{P. Kumar \& B. Zhang(2015)Kumar \& Zhang}]{Kumar_2015}
Kumar, P., \& Zhang, B. 2015, \bibinfo{title}{The physics of gamma-ray bursts and relativistic jets,} Physics Reports, 561, 1–109, \dodoi{10.1016/j.physrep.2014.09.008}

\bibitem[{Y. {Kusafuka} {et~al.}(2025){Kusafuka}, {Matsuoka}, \& {Sawada}}]{2025arXiv250313291K}
{Kusafuka}, Y., {Matsuoka}, T., \& {Sawada}, R. 2025, \bibinfo{title}{{Contribution to the Radio Light Curves of a Core-Collapse Supernova due to the Off-axis Gamma-Ray Burst Afterglow},} arXiv e-prints, arXiv:2503.13291, \dodoi{10.48550/arXiv.2503.13291}

\bibitem[{M. {Lacy} {et~al.}(2020){Lacy}, {Baum}, {Chandler}, {Chatterjee}, {Clarke}, {Deustua}, {English}, {Farnes}, {Gaensler}, {Gugliucci}, {Hallinan}, {Kent}, {Kimball}, {Law}, {Lazio}, {Marvil}, {Mao}, {Medlin}, {Mooley}, {Murphy}, {Myers}, {Osten}, {Richards}, {Rosolowsky}, {Rudnick}, {Schinzel}, {Sivakoff}, {Sjouwerman}, {Taylor}, {White}, {Wrobel}, {Andernach}, {Beasley}, {Berger}, {Bhatnager}, {Birkinshaw}, {Bower}, {Brandt}, {Brown}, {Burke-Spolaor}, {Butler}, {Comerford}, {Demorest}, {Fu}, {Giacintucci}, {Golap}, {G{\"u}th}, {Hales}, {Hiriart}, {Hodge}, {Horesh}, {Ivezi{\'c}}, {Jarvis}, {Kamble}, {Kassim}, {Liu}, {Loinard}, {Lyons}, {Masters}, {Mezcua}, {Moellenbrock}, {Mroczkowski}, {Nyland}, {O'Dea}, {O'Sullivan}, {Peters}, {Radford}, {Rao}, {Robnett}, {Salcido}, {Shen}, {Sobotka}, {Witz}, {Vaccari}, {van Weeren}, {Vargas}, {Williams}, \& {Yoon}}]{2020PASP..132c5001L}
{Lacy}, M., {Baum}, S.~A., {Chandler}, C.~J., {et~al.} 2020, \bibinfo{title}{{The Karl G. Jansky Very Large Array Sky Survey (VLASS). Science Case and Survey Design},} \pasp, 132, 035001, \dodoi{10.1088/1538-3873/ab63eb}

\bibitem[{T. {Laskar} {et~al.}(2014){Laskar}, {Berger}, {Tanvir}, {Zauderer}, {Margutti}, {Levan}, {Perley}, {Fong}, {Wiersema}, {Menten}, \& {Hrudkova}}]{2014ApJ...781....1L}
{Laskar}, T., {Berger}, E., {Tanvir}, N., {et~al.} 2014, \bibinfo{title}{{GRB 120521C at z \raisebox{-0.5ex}\textasciitilde 6 and the Properties of High-redshift {\ensuremath{\gamma}}-Ray Bursts},} \apj, 781, 1, \dodoi{10.1088/0004-637X/781/1/1}

\bibitem[{T. {Laskar} {et~al.}(2018){Laskar}, {Alexander}, {Berger}, {Guidorzi}, {Margutti}, {Fong}, {Kilpatrick}, {Milne}, {Drout}, {Mundell}, {Kobayashi}, {Lunnan}, {Barniol Duran}, {Menten}, {Ioka}, \& {Williams}}]{2018ApJ...862...94L}
{Laskar}, T., {Alexander}, K.~D., {Berger}, E., {et~al.} 2018, \bibinfo{title}{{First ALMA Light Curve Constrains Refreshed Reverse Shocks and Jet Magnetization in GRB 161219B},} \apj, 862, 94, \dodoi{10.3847/1538-4357/aacbcc}

\bibitem[{C.~J. {Law} {et~al.}(2018){Law}, {Gaensler}, {Metzger}, {Ofek}, \& {Sironi}}]{2018ApJ...866L..22L}
{Law}, C.~J., {Gaensler}, B.~M., {Metzger}, B.~D., {Ofek}, E.~O., \& {Sironi}, L. 2018, \bibinfo{title}{{Discovery of the Luminous, Decades-long, Extragalactic Radio Transient FIRST J141918.9+394036},} \apjl, 866, L22, \dodoi{10.3847/2041-8213/aae5f3}

\bibitem[{N.~M. {Law} {et~al.}(2009){Law}, {Kulkarni}, {Dekany}, {Ofek}, {Quimby}, {Nugent}, {Surace}, {Grillmair}, {Bloom}, {Kasliwal}, {Bildsten}, {Brown}, {Cenko}, {Ciardi}, {Croner}, {Djorgovski}, {van Eyken}, {Filippenko}, {Fox}, {Gal-Yam}, {Hale}, {Hamam}, {Helou}, {Henning}, {Howell}, {Jacobsen}, {Laher}, {Mattingly}, {McKenna}, {Pickles}, {Poznanski}, {Rahmer}, {Rau}, {Rosing}, {Shara}, {Smith}, {Starr}, {Sullivan}, {Velur}, {Walters}, \& {Zolkower}}]{2009PASP..121.1395L}
{Law}, N.~M., {Kulkarni}, S.~R., {Dekany}, R.~G., {et~al.} 2009, \bibinfo{title}{{The Palomar Transient Factory: System Overview, Performance, and First Results},} \pasp, 121, 1395, \dodoi{10.1086/648598}

\bibitem[{E. {Liang} {et~al.}(2007){Liang}, {Zhang}, {Virgili}, \& {Dai}}]{2007ApJ...662.1111L}
{Liang}, E., {Zhang}, B., {Virgili}, F., \& {Dai}, Z.~G. 2007, \bibinfo{title}{{Low-Luminosity Gamma-Ray Bursts as a Unique Population: Luminosity Function, Local Rate, and Beaming Factor},} \apj, 662, 1111, \dodoi{10.1086/517959}

\bibitem[{X. {Ma} {et~al.}(2025{\natexlab{a}}){Ma}, {Wang}, {Mo}, {Howell}, {Pellegrino}, {Zhang}, {Yan}, {Arcavi}, {Chen}, {Farah}, {Padilla Gonzalez}, {Guo}, {Hiramatsu}, {Li}, {Lin}, {Liu}, {McCully}, {Newsome}, {Sai}, {Terreran}, {Xiang}, {Zhang}, \& {Zhang}}]{2025arXiv250404393M}
{Ma}, X., {Wang}, X., {Mo}, J., {et~al.} 2025{\natexlab{a}}, \bibinfo{title}{{Supernovae at Distances < 40 Mpc: I.Catalogues and fractions of Supernovae in a Complete Sample},} arXiv e-prints, arXiv:2504.04393, \dodoi{10.48550/arXiv.2504.04393}

\bibitem[{X. {Ma} {et~al.}(2025{\natexlab{b}}){Ma}, {Wang}, {Mo}, {Howell}, {Pellegrino}, {Zhang}, {Wu}, {Yan}, {Liu}, {Arcavi}, {Chen}, {Farah}, {Padilla Gonzalez}, {Guo}, {Hiramatsu}, {Li}, {Lin}, {Liu}, {McCully}, {Newsome}, {Sai}, {Terreran}, {Xiang}, \& {Zhang}}]{2025arXiv250404507M}
{Ma}, X., {Wang}, X., {Mo}, J., {et~al.} 2025{\natexlab{b}}, \bibinfo{title}{{Supernovae at Distances < 40 Mpc: II. Supernova Rate in the Local Universe},} arXiv e-prints, arXiv:2504.04507, \dodoi{10.48550/arXiv.2504.04507}

\bibitem[{R. {Margutti} {et~al.}(2013){Margutti}, {Soderberg}, {Wieringa}, {Edwards}, {Chevalier}, {Morsony}, {Barniol Duran}, {Sironi}, {Zauderer}, {Milisavljevic}, {Kamble}, \& {Pian}}]{2013ApJ...778...18M}
{Margutti}, R., {Soderberg}, A.~M., {Wieringa}, M.~H., {et~al.} 2013, \bibinfo{title}{{The Signature of the Central Engine in the Weakest Relativistic Explosions: GRB 100316D},} \apj, 778, 18, \dodoi{10.1088/0004-637X/778/1/18}

\bibitem[{M. {Marongiu} {et~al.}(2019){Marongiu}, {Guidorzi}, {Margutti}, {Coppejans}, {Martone}, \& {Kamble}}]{2019ApJ...879...89M}
{Marongiu}, M., {Guidorzi}, C., {Margutti}, R., {et~al.} 2019, \bibinfo{title}{{Constraints on the Environment and Energetics of the Broad-line Ic SN2014ad from Deep Radio and X-Ray Observations},} \apj, 879, 89, \dodoi{10.3847/1538-4357/ab25ef}

\bibitem[{P.~A. {Mazzali} {et~al.}(2021){Mazzali}, {Pian}, {Bufano}, \& {Ashall}}]{2021MNRAS.505.4106M}
{Mazzali}, P.~A., {Pian}, E., {Bufano}, F., \& {Ashall}, C. 2021, \bibinfo{title}{{Modelling of SN 2013dx associated with the low-redshift GRB130702A points to diversity in GRB/SN properties},} \mnras, 505, 4106, \dodoi{10.1093/mnras/stab1594}

\bibitem[{D. {McConnell} {et~al.}(2020){McConnell}, {Hale}, {Lenc}, {Banfield}, {Heald}, {Hotan}, {Leung}, {Moss}, {Murphy}, {O'Brien}, {Pritchard}, {Raja}, {Sadler}, {Stewart}, {Thomson}, {Whiting}, {Allison}, {Amy}, {Anderson}, {Ball}, {Bannister}, {Bell}, {Bock}, {Bolton}, {Bunton}, {Chippendale}, {Collier}, {Cooray}, {Cornwell}, {Diamond}, {Edwards}, {Gupta}, {Hayman}, {Heywood}, {Jackson}, {Koribalski}, {Lee-Waddell}, {McClure-Griffiths}, {Ng}, {Norris}, {Phillips}, {Reynolds}, {Roxby}, {Schinckel}, {Shields}, {Tremblay}, {Tzioumis}, {Voronkov}, \& {Westmeier}}]{2020PASA...37...48M}
{McConnell}, D., {Hale}, C.~L., {Lenc}, E., {et~al.} 2020, \bibinfo{title}{{The Rapid ASKAP Continuum Survey I: Design and first results},} \pasa, 37, e048, \dodoi{10.1017/pasa.2020.41}

\bibitem[{W. Mckinney(2010)Mckinney}]{pandas}
Mckinney, W. 2010, \bibinfo{title}{Data Structures for Statistical Computing in Python,} Proceedings of the 9th Python in Science Conference

\bibitem[{J.~P. {McMullin} {et~al.}(2007){McMullin}, {Waters}, {Schiebel}, {Young}, \& {Golap}}]{2007ASPC..376..127M}
{McMullin}, J.~P., {Waters}, B., {Schiebel}, D., {Young}, W., \& {Golap}, K. 2007, \bibinfo{title}{{CASA Architecture and Applications},} in Astronomical Society of the Pacific Conference Series, Vol. 376, Astronomical Data Analysis Software and Systems XVI, ed. R.~A. {Shaw}, F.~{Hill}, \& D.~J. {Bell}, 127

\bibitem[{C. {Meegan} {et~al.}(2009){Meegan}, {Lichti}, {Bhat}, {Bissaldi}, {Briggs}, {Connaughton}, {Diehl}, {Fishman}, {Greiner}, {Hoover}, {van der Horst}, {von Kienlin}, {Kippen}, {Kouveliotou}, {McBreen}, {Paciesas}, {Preece}, {Steinle}, {Wallace}, {Wilson}, \& {Wilson-Hodge}}]{Meegan2009}
{Meegan}, C., {Lichti}, G., {Bhat}, P.~N., {et~al.} 2009, \bibinfo{title}{{The Fermi Gamma-ray Burst Monitor},} \apj, 702, 791, \dodoi{10.1088/0004-637X/702/1/791}

\bibitem[{A. {Melandri} {et~al.}(2014){Melandri}, {Pian}, {D'Elia}, {D'Avanzo}, {Della Valle}, {Mazzali}, {Tagliaferri}, {Cano}, {Levan}, {M{\o}oller}, {Amati}, {Bernardini}, {Bersier}, {Bufano}, {Campana}, {Castro-Tirado}, {Covino}, {Ghirlanda}, {Hurley}, {Malesani}, {Masetti}, {Palazzi}, {Piranomonte}, {Rossi}, {Salvaterra}, {Starling}, {Tanaka}, {Tanvir}, \& {Vergani}}]{2014A&A...567A..29M}
{Melandri}, A., {Pian}, E., {D'Elia}, V., {et~al.} 2014, \bibinfo{title}{{Diversity of gamma-ray burst energetics vs. supernova homogeneity: SN 2013cq associated with GRB 130427A},} \aap, 567, A29, \dodoi{10.1051/0004-6361/201423572}

\bibitem[{P. {M{\'e}sz{\'a}ros}(2006){M{\'e}sz{\'a}ros}}]{Meszaros2006}
{M{\'e}sz{\'a}ros}, P. 2006, \bibinfo{title}{{Gamma-ray bursts},} Reports on Progress in Physics, 69, 2259, \dodoi{10.1088/0034-4885/69/8/R01}

\bibitem[{P. {M{\'e}sz{\'a}ros} \& E. {Waxman}(2001){M{\'e}sz{\'a}ros} \& {Waxman}}]{2001PhRvL..87q1102M}
{M{\'e}sz{\'a}ros}, P., \& {Waxman}, E. 2001, \bibinfo{title}{{TeV Neutrinos from Successful and Choked Gamma-Ray Bursts},} \prl, 87, 171102, \dodoi{10.1103/PhysRevLett.87.171102}

\bibitem[{A. Mizuta \& K. Ioka(2013)Mizuta \& Ioka}]{Mizuta+2013}
Mizuta, A., \& Ioka, K. 2013, \bibinfo{title}{OPENING ANGLES OF COLLAPSAR JETS,} \apj, 777, 162, \dodoi{10.1088/0004-637X/777/2/162}

\bibitem[{M. {Modjaz} {et~al.}(2019){Modjaz}, {Guti{\'e}rrez}, \& {Arcavi}}]{2019NatAs...3..717M}
{Modjaz}, M., {Guti{\'e}rrez}, C.~P., \& {Arcavi}, I. 2019, \bibinfo{title}{{New regimes in the observation of core-collapse supernovae},} Nature Astronomy, 3, 717, \dodoi{10.1038/s41550-019-0856-2}

\bibitem[{M. {Modjaz} {et~al.}(2016){Modjaz}, {Liu}, {Bianco}, \& {Graur}}]{Modjaz2016}
{Modjaz}, M., {Liu}, Y.~Q., {Bianco}, F.~B., \& {Graur}, O. 2016, \bibinfo{title}{{The Spectral SN-GRB Connection: Systematic Spectral Comparisons between Type Ic Supernovae and Broad-lined Type Ic Supernovae with and without Gamma-Ray Bursts},} \apj, 832, 108, \dodoi{10.3847/0004-637X/832/2/108}

\bibitem[{M. {Modjaz} {et~al.}(2020){Modjaz}, {Bianco}, {Siwek}, {Huang}, {Perley}, {Fierroz}, {Liu}, {Arcavi}, {Gal-Yam}, {Filippenko}, {Blagorodnova}, {Cenko}, {Kasliwal}, {Kulkarni}, {Schulze}, {Taggart}, \& {Zheng}}]{2020ApJ...892..153M}
{Modjaz}, M., {Bianco}, F.~B., {Siwek}, M., {et~al.} 2020, \bibinfo{title}{{Host Galaxies of Type Ic and Broad-lined Type Ic Supernovae from the Palomar Transient Factory: Implications for Jet Production},} \apj, 892, 153, \dodoi{10.3847/1538-4357/ab4185}

\bibitem[{K.~P. {Mooley} {et~al.}(2016){Mooley}, {Hallinan}, {Bourke}, {Horesh}, {Myers}, {Frail}, {Kulkarni}, {Levitan}, {Kasliwal}, {Cenko}, {Cao}, {Bellm}, \& {Laher}}]{2016ApJ...818..105M}
{Mooley}, K.~P., {Hallinan}, G., {Bourke}, S., {et~al.} 2016, \bibinfo{title}{{The Caltech-NRAO Stripe 82 Survey (CNSS). I. The Pilot Radio Transient Survey In 50 deg$^{2}$},} \apj, 818, 105, \dodoi{10.3847/0004-637X/818/2/105}

\bibitem[{K.~P. {Mooley} {et~al.}(2022){Mooley}, {Margalit}, {Law}, {Perley}, {Deller}, {Lazio}, {Bietenholz}, {Shimwell}, {Intema}, {Gaensler}, {Metzger}, {Dong}, {Hallinan}, {Ofek}, \& {Sironi}}]{2022ApJ...924...16M}
{Mooley}, K.~P., {Margalit}, B., {Law}, C.~J., {et~al.} 2022, \bibinfo{title}{{Late-time Evolution and Modeling of the Off-axis Gamma-Ray Burst Candidate FIRST J141918.9+394036},} \apj, 924, 16, \dodoi{10.3847/1538-4357/ac3330}

\bibitem[{M.~J. {Moss} {et~al.}(2023){Moss}, {Mochkovitch}, {Daigne}, {Beniamini}, \& {Guiriec}}]{2023MNRAS.525.5224M}
{Moss}, M.~J., {Mochkovitch}, R., {Daigne}, F., {Beniamini}, P., \& {Guiriec}, S. 2023, \bibinfo{title}{{The signature of refreshed shocks in the afterglow of GRB 030329},} \mnras, 525, 5224, \dodoi{10.1093/mnras/stad2594}

\bibitem[{E. {Nakar}(2015){Nakar}}]{2015ApJ...807..172N}
{Nakar}, E. 2015, \bibinfo{title}{{A Unified Picture for Low-luminosity and Long Gamma-Ray Bursts Based on the Extended Progenitor of llGRB 060218/SN 2006aj},} \apj, 807, 172, \dodoi{10.1088/0004-637X/807/2/172}

\bibitem[{N.~T. {Palliyaguru} {et~al.}(2021){Palliyaguru}, {Corsi}, {P{\'e}rez-Torres}, {Varenius}, \& {Van Eerten}}]{2021ApJ...910...16P}
{Palliyaguru}, N.~T., {Corsi}, A., {P{\'e}rez-Torres}, M., {Varenius}, E., \& {Van Eerten}, H. 2021, \bibinfo{title}{{VLBI Observations of Supernova PTF11qcj: Direct Constraints on the Size of the Radio Ejecta},} \apj, 910, 16, \dodoi{10.3847/1538-4357/abe1c9}

\bibitem[{N.~T. {Palliyaguru} {et~al.}(2019){Palliyaguru}, {Corsi}, {Frail}, {Vink{\'o}}, {Wheeler}, {Gal-Yam}, {Cenko}, {Kulkarni}, \& {Kasliwal}}]{2019ApJ...872..201P}
{Palliyaguru}, N.~T., {Corsi}, A., {Frail}, D.~A., {et~al.} 2019, \bibinfo{title}{{The Double-peaked Radio Light Curve of Supernova PTF11qcj},} \apj, 872, 201, \dodoi{10.3847/1538-4357/aaf64d}

\bibitem[{V.~D. {Pal'shin} {et~al.}(2013){Pal'shin}, {Hurley}, {Svinkin}, {Aptekar}, {Golenetskii}, {Frederiks}, {Mazets}, {Oleynik}, {Ulanov}, {Cline}, {Mitrofanov}, {Golovin}, {Kozyrev}, {Litvak}, {Sanin}, {Boynton}, {Fellows}, {Harshman}, {Trombka}, {McClanahan}, {Starr}, {Goldsten}, {Gold}, {Rau}, {von Kienlin}, {Savchenko}, {Smith}, {Hajdas}, {Barthelmy}, {Cummings}, {Gehrels}, {Krimm}, {Palmer}, {Yamaoka}, {Ohno}, {Fukazawa}, {Hanabata}, {Takahashi}, {Tashiro}, {Terada}, {Murakami}, {Makishima}, {Briggs}, {Kippen}, {Kouveliotou}, {Meegan}, {Fishman}, {Connaughton}, {Bo{\"e}r}, {Guidorzi}, {Frontera}, {Montanari}, {Rossi}, {Feroci}, {Amati}, {Nicastro}, {Orlandini}, {Del Monte}, {Costa}, {Donnarumma}, {Evangelista}, {Lapshov}, {Lazzarotto}, {Pacciani}, {Rapisarda}, {Soffitta}, {Di Cocco}, {Fuschino}, {Galli}, {Labanti}, {Marisaldi}, {Atteia}, {Vanderspek}, \& {Ricker}}]{Palshin2013}
{Pal'shin}, V.~D., {Hurley}, K., {Svinkin}, D.~S., {et~al.} 2013, \bibinfo{title}{{Interplanetary Network Localizations of Konus Short Gamma-Ray Bursts},} \apjs, 207, 38, \dodoi{10.1088/0067-0049/207/2/38}

\bibitem[{A. {Panaitescu} \& P. {Kumar}(2002){Panaitescu} \& {Kumar}}]{2002ApJ...571..779P}
{Panaitescu}, A., \& {Kumar}, P. 2002, \bibinfo{title}{{Properties of Relativistic Jets in Gamma-Ray Burst Afterglows},} \apj, 571, 779, \dodoi{10.1086/340094}

\bibitem[{D.~A. {Perley} {et~al.}(2020){Perley}, {Fremling}, {Sollerman}, {Miller}, {Dahiwale}, {Sharma}, {Bellm}, {Biswas}, {Brink}, {Bruch}, {De}, {Dekany}, {Drake}, {Duev}, {Filippenko}, {Gal-Yam}, {Goobar}, {Graham}, {Graham}, {Ho}, {Irani}, {Kasliwal}, {Kim}, {Kulkarni}, {Mahabal}, {Masci}, {Modak}, {Neill}, {Nordin}, {Riddle}, {Soumagnac}, {Strotjohann}, {Schulze}, {Taggart}, {Tzanidakis}, {Walters}, \& {Yan}}]{2020ApJ...904...35P}
{Perley}, D.~A., {Fremling}, C., {Sollerman}, J., {et~al.} 2020, \bibinfo{title}{{The Zwicky Transient Facility Bright Transient Survey. II. A Public Statistical Sample for Exploring Supernova Demographics},} \apj, 904, 35, \dodoi{10.3847/1538-4357/abbd98}

\bibitem[{R.~A. {Perley} \& B.~J. {Butler}(2017){Perley} \& {Butler}}]{2017ApJS..230....7P}
{Perley}, R.~A., \& {Butler}, B.~J. 2017, \bibinfo{title}{{An Accurate Flux Density Scale from 50 MHz to 50 GHz},} \apjs, 230, 7, \dodoi{10.3847/1538-4365/aa6df9}

\bibitem[{T. Piran(2005)Piran}]{Piran_2005}
Piran, T. 2005, \bibinfo{title}{The physics of gamma-ray bursts,} Reviews of Modern Physics, 76, 1143–1210, \dodoi{10.1103/revmodphys.76.1143}

\bibitem[{ {Planck Collaboration} {et~al.}(2020){Planck Collaboration}, {Aghanim}, {Akrami}, {Ashdown}, {Aumont}, {Baccigalupi}, {Ballardini}, {Banday}, {Barreiro}, {Bartolo}, {Basak}, {Battye}, {Benabed}, {Bernard}, {Bersanelli}, {Bielewicz}, {Bock}, {Bond}, {Borrill}, {Bouchet}, {Boulanger}, {Bucher}, {Burigana}, {Butler}, {Calabrese}, {Cardoso}, {Carron}, {Challinor}, {Chiang}, {Chluba}, {Colombo}, {Combet}, {Contreras}, {Crill}, {Cuttaia}, {de Bernardis}, {de Zotti}, {Delabrouille}, {Delouis}, {Di Valentino}, {Diego}, {Dor{\'e}}, {Douspis}, {Ducout}, {Dupac}, {Dusini}, {Efstathiou}, {Elsner}, {En{\ss}lin}, {Eriksen}, {Fantaye}, {Farhang}, {Fergusson}, {Fernandez-Cobos}, {Finelli}, {Forastieri}, {Frailis}, {Fraisse}, {Franceschi}, {Frolov}, {Galeotta}, {Galli}, {Ganga}, {G{\'e}nova-Santos}, {Gerbino}, {Ghosh}, {Gonz{\'a}lez-Nuevo}, {G{\'o}rski}, {Gratton}, {Gruppuso}, {Gudmundsson}, {Hamann}, {Handley}, {Hansen}, {Herranz}, {Hildebrandt}, {Hivon}, {Huang}, {Jaffe}, {Jones}, {Karakci}, {Keih{\"a}nen},
  {Keskitalo}, {Kiiveri}, {Kim}, {Kisner}, {Knox}, {Krachmalnicoff}, {Kunz}, {Kurki-Suonio}, {Lagache}, {Lamarre}, {Lasenby}, {Lattanzi}, {Lawrence}, {Le Jeune}, {Lemos}, {Lesgourgues}, {Levrier}, {Lewis}, {Liguori}, {Lilje}, {Lilley}, {Lindholm}, {L{\'o}pez-Caniego}, {Lubin}, {Ma}, {Mac{\'\i}as-P{\'e}rez}, {Maggio}, {Maino}, {Mandolesi}, {Mangilli}, {Marcos-Caballero}, {Maris}, {Martin}, {Martinelli}, {Mart{\'\i}nez-Gonz{\'a}lez}, {Matarrese}, {Mauri}, {McEwen}, {Meinhold}, {Melchiorri}, {Mennella}, {Migliaccio}, {Millea}, {Mitra}, {Miville-Desch{\^e}nes}, {Molinari}, {Montier}, {Morgante}, {Moss}, {Natoli}, {N{\o}rgaard-Nielsen}, {Pagano}, {Paoletti}, {Partridge}, {Patanchon}, {Peiris}, {Perrotta}, {Pettorino}, {Piacentini}, {Polastri}, {Polenta}, {Puget}, {Rachen}, {Reinecke}, {Remazeilles}, {Renzi}, {Rocha}, {Rosset}, {Roudier}, {Rubi{\~n}o-Mart{\'\i}n}, {Ruiz-Granados}, {Salvati}, {Sandri}, {Savelainen}, {Scott}, {Shellard}, {Sirignano}, {Sirri}, {Spencer}, {Sunyaev}, {Suur-Uski}, {Tauber}, {Tavagnacco},
  {Tenti}, {Toffolatti}, {Tomasi}, {Trombetti}, {Valenziano}, {Valiviita}, {Van Tent}, {Vibert}, {Vielva}, {Villa}, {Vittorio}, {Wandelt}, {Wehus}, {White}, {White}, {Zacchei}, \& {Zonca}}]{2020A&A...641A...6P}
{Planck Collaboration}, {Aghanim}, N., {Akrami}, Y., {et~al.} 2020, \bibinfo{title}{{Planck 2018 results. VI. Cosmological parameters},} \aap, 641, A6, \dodoi{10.1051/0004-6361/201833910}

\bibitem[{J.~E. {Rhoads}(1997){Rhoads}}]{Rhoads1997}
{Rhoads}, J.~E. 1997, \bibinfo{title}{{How to Tell a Jet from a Balloon: A Proposed Test for Beaming in Gamma-Ray Bursts},} \apjl, 487, L1, \dodoi{10.1086/310876}

\bibitem[{G. {Ryan} {et~al.}(2015){Ryan}, {van Eerten}, {MacFadyen}, \& {Zhang}}]{2015ApJ...799....3R}
{Ryan}, G., {van Eerten}, H., {MacFadyen}, A., \& {Zhang}, B.-B. 2015, \bibinfo{title}{{Gamma-Ray Bursts are Observed Off-axis},} \apj, 799, 3, \dodoi{10.1088/0004-637X/799/1/3}

\bibitem[{R. {Sari} {et~al.}(1999){Sari}, {Piran}, \& {Halpern}}]{Sari1999}
{Sari}, R., {Piran}, T., \& {Halpern}, J.~P. 1999, \bibinfo{title}{{Jets in Gamma-Ray Bursts},} \apjl, 519, L17, \dodoi{10.1086/312109}

\bibitem[{R. {Sari} {et~al.}(1998){Sari}, {Piran}, \& {Narayan}}]{1998ApJ...497L..17S}
{Sari}, R., {Piran}, T., \& {Narayan}, R. 1998, \bibinfo{title}{{Spectra and Light Curves of Gamma-Ray Burst Afterglows},} \apjl, 497, L17, \dodoi{10.1086/311269}

\bibitem[{G. {Schroeder} {et~al.}(2022){Schroeder}, {Laskar}, {Fong}, {Nugent}, {Berger}, {Chornock}, {Alexander}, {Andrews}, {Bussmann}, {Castro-Tirado}, {Goyal}, {Kilpatrick}, {Lally}, {Miller}, {Milne}, {Paterson}, {Escorial}, {Stroh}, {Terreran}, \& {Zauderer}}]{2022ApJ...940...53S}
{Schroeder}, G., {Laskar}, T., {Fong}, W.-f., {et~al.} 2022, \bibinfo{title}{{A Radio-selected Population of Dark, Long Gamma-Ray Bursts: Comparison to the Long Gamma-Ray Burst Population and Implications for Host Dust Distributions},} \apj, 940, 53, \dodoi{10.3847/1538-4357/ac8feb}

\bibitem[{T.~W. {Shimwell} {et~al.}(2022){Shimwell}, {Hardcastle}, {Tasse}, {Best}, {R{\"o}ttgering}, {Williams}, {Botteon}, {Drabent}, {Mechev}, {Shulevski}, {van Weeren}, {Bester}, {Br{\"u}ggen}, {Brunetti}, {Callingham}, {Chy{\.z}y}, {Conway}, {Dijkema}, {Duncan}, {de Gasperin}, {Hale}, {Haverkorn}, {Hugo}, {Jackson}, {Mevius}, {Miley}, {Morabito}, {Morganti}, {Offringa}, {Oonk}, {Rafferty}, {Sabater}, {Smith}, {Schwarz}, {Smirnov}, {O'Sullivan}, {Vedantham}, {White}, {Albert}, {Alegre}, {Asabere}, {Bacon}, {Bonafede}, {Bonnassieux}, {Brienza}, {Bilicki}, {Bonato}, {Calistro Rivera}, {Cassano}, {Cochrane}, {Croston}, {Cuciti}, {Dallacasa}, {Danezi}, {Dettmar}, {Di Gennaro}, {Edler}, {En{\ss}lin}, {Emig}, {Franzen}, {Garc{\'\i}a-Vergara}, {Grange}, {G{\"u}rkan}, {Hajduk}, {Heald}, {Heesen}, {Hoang}, {Hoeft}, {Horellou}, {Iacobelli}, {Jamrozy}, {Jeli{\'c}}, {Kondapally}, {Kukreti}, {Kunert-Bajraszewska}, {Magliocchetti}, {Mahatma}, {Ma{\l}ek}, {Mandal}, {Massaro}, {Meyer-Zhao}, {Mingo}, {Mostert}, {Nair},
  {Nakoneczny}, {Nikiel-Wroczy{\'n}ski}, {Orr{\'u}}, {Pajdosz-{\'S}mierciak}, {Pasini}, {Prandoni}, {van Piggelen}, {Rajpurohit}, {Retana-Montenegro}, {Riseley}, {Rowlinson}, {Saxena}, {Schrijvers}, {Sweijen}, {Siewert}, {Timmerman}, {Vaccari}, {Vink}, {West}, {Wo{\l}owska}, {Zhang}, \& {Zheng}}]{2022A&A...659A...1S}
{Shimwell}, T.~W., {Hardcastle}, M.~J., {Tasse}, C., {et~al.} 2022, \bibinfo{title}{{The LOFAR Two-metre Sky Survey. V. Second data release},} \aap, 659, A1, \dodoi{10.1051/0004-6361/202142484}

\bibitem[{V. {Smol{\v{c}}i{\'c}} {et~al.}(2017){Smol{\v{c}}i{\'c}}, {Delvecchio}, {Zamorani}, {Baran}, {Novak}, {Delhaize}, {Schinnerer}, {Berta}, {Bondi}, {Ciliegi}, {Capak}, {Civano}, {Karim}, {Le Fevre}, {Ilbert}, {Laigle}, {Marchesi}, {McCracken}, {Tasca}, {Salvato}, \& {Vardoulaki}}]{2017A&A...602A...2S}
{Smol{\v{c}}i{\'c}}, V., {Delvecchio}, I., {Zamorani}, G., {et~al.} 2017, \bibinfo{title}{{The VLA-COSMOS 3 GHz Large Project: Multiwavelength counterparts and the composition of the faint radio population},} \aap, 602, A2, \dodoi{10.1051/0004-6361/201630223}

\bibitem[{A.~M. Soderberg {et~al.}(2006)Soderberg, Nakar, Berger, \& Kulkarni}]{2006ApJ...638..930S}
Soderberg, A.~M., Nakar, E., Berger, E., \& Kulkarni, S.~R. 2006, \bibinfo{title}{Late-{Time} {Radio} {Observations} of 68 {Type} {Ibc} {Supernovae}: {Strong} {Constraints} on {Off}-{Axis} {Gamma}-{Ray} {Bursts},} \apj, 638, 930, \dodoi{10.1086/499121}

\bibitem[{A.~M. {Soderberg} {et~al.}(2004){Soderberg}, {Kulkarni}, {Berger}, {Fox}, {Sako}, {Frail}, {Gal-Yam}, {Moon}, {Cenko}, {Yost}, {Phillips}, {Persson}, {Freedman}, {Wyatt}, {Jayawardhana}, \& {Paulson}}]{2004Natur.430..648S}
{Soderberg}, A.~M., {Kulkarni}, S.~R., {Berger}, E., {et~al.} 2004, \bibinfo{title}{{The sub-energetic {\ensuremath{\gamma}}-ray burst GRB 031203 as a cosmic analogue to the nearby GRB 980425},} \nat, 430, 648, \dodoi{10.1038/nature02757}

\bibitem[{A.~M. {Soderberg} {et~al.}(2006){Soderberg}, {Kulkarni}, {Nakar}, {Berger}, {Cameron}, {Fox}, {Frail}, {Gal-Yam}, {Sari}, {Cenko}, {Kasliwal}, {Chevalier}, {Piran}, {Price}, {Schmidt}, {Pooley}, {Moon}, {Penprase}, {Ofek}, {Rau}, {Gehrels}, {Nousek}, {Burrows}, {Persson}, \& {McCarthy}}]{2006Natur.442.1014S}
{Soderberg}, A.~M., {Kulkarni}, S.~R., {Nakar}, E., {et~al.} 2006, \bibinfo{title}{{Relativistic ejecta from X-ray flash XRF 060218 and the rate of cosmic explosions},} \nat, 442, 1014, \dodoi{10.1038/nature05087}

\bibitem[{A.~M. {Soderberg} {et~al.}(2010){Soderberg}, {Chakraborti}, {Pignata}, {Chevalier}, {Chandra}, {Ray}, {Wieringa}, {Copete}, {Chaplin}, {Connaughton}, {Barthelmy}, {Bietenholz}, {Chugai}, {Stritzinger}, {Hamuy}, {Fransson}, {Fox}, {Levesque}, {Grindlay}, {Challis}, {Foley}, {Kirshner}, {Milne}, \& {Torres}}]{2010Natur.463..513S}
{Soderberg}, A.~M., {Chakraborti}, S., {Pignata}, G., {et~al.} 2010, \bibinfo{title}{{A relativistic type Ibc supernova without a detected {\ensuremath{\gamma}}-ray burst},} \nat, 463, 513, \dodoi{10.1038/nature08714}

\bibitem[{G.~P. {Srinivasaragavan} {et~al.}(2024){Srinivasaragavan}, {Yang}, {Anand}, {Sollerman}, {Ho}, {Corsi}, {Cenko}, {Perley}, {Schulze}, {Sanchez-Fleming}, {Pope}, {Sarin}, {Omand}, {Das}, {Fremling}, {Andreoni}, {Bruch}, {Burdge}, {De}, {Gal-Yam}, {Gangopadhyay}, {Graham}, {Jencson}, {Karambelkar}, {Kasliwal}, {Kulkarni}, {Martikainen}, {Sharma}, {Tzanidakis}, {Yan}, {Yao}, {Bellm}, {Groom}, {Masci}, {Nir}, {Purdum}, {Smith}, \& {Sravan}}]{Srinivasaragavan2024}
{Srinivasaragavan}, G.~P., {Yang}, S., {Anand}, S., {et~al.} 2024, \bibinfo{title}{{Optical and Radio Analysis of Systematically Classified Broad-lined Type Ic Supernovae from the Zwicky Transient Facility},} \apj, 976, 71, \dodoi{10.3847/1538-4357/ad7fde}

\bibitem[{M.~C. Stroh {et~al.}(2021)Stroh, Terreran, Coppejans, Bright, Margutti, Bietenholz, De~Colle, DeMarchi, Duran, Milisavljevic, Murase, Paterson, \& Williams}]{2021ApJ...923L..24S}
Stroh, M.~C., Terreran, G., Coppejans, D.~L., {et~al.} 2021, \bibinfo{title}{Luminous {Late}-time {Radio} {Emission} from {Supernovae} {Detected} by the {Karl} {G}. {Jansky} {Very} {Large} {Array} {Sky} {Survey} ({VLASS}),} \apj, 923, L24, \dodoi{10.3847/2041-8213/ac375e}

\bibitem[{F. {Taddia} {et~al.}(2019){Taddia}, {Sollerman}, {Fremling}, {Barbarino}, {Karamehmetoglu}, {Arcavi}, {Cenko}, {Filippenko}, {Gal-Yam}, {Hiramatsu}, {Hosseinzadeh}, {Howell}, {Kulkarni}, {Laher}, {Lunnan}, {Masci}, {Nugent}, {Nyholm}, {Perley}, {Quimby}, \& {Silverman}}]{2019A&A...621A..71T}
{Taddia}, F., {Sollerman}, J., {Fremling}, C., {et~al.} 2019, \bibinfo{title}{{Analysis of broad-lined Type Ic supernovae from the (intermediate) Palomar Transient Factory},} \aap, 621, A71, \dodoi{10.1051/0004-6361/201834429}

\bibitem[{N. {Thyagarajan} {et~al.}(2011){Thyagarajan}, {Helfand}, {White}, \& {Becker}}]{2011ApJ...742...49T}
{Thyagarajan}, N., {Helfand}, D.~J., {White}, R.~L., \& {Becker}, R.~H. 2011, \bibinfo{title}{{Variable and Transient Radio Sources in the FIRST Survey},} \apj, 742, 49, \dodoi{10.1088/0004-637X/742/1/49}

\bibitem[{A.~J. {van der Horst} {et~al.}(2008){van der Horst}, {Kamble}, {Resmi}, {Wijers}, {Bhattacharya}, {Scheers}, {Rol}, {Strom}, {Kouveliotou}, {Oosterloo}, \& {Ishwara-Chandra}}]{2008A&A...480...35V}
{van der Horst}, A.~J., {Kamble}, A., {Resmi}, L., {et~al.} 2008, \bibinfo{title}{{Detailed study of the GRB 030329 radio afterglow deep into the non-relativistic phase},} \aap, 480, 35, \dodoi{10.1051/0004-6361:20078051}

\bibitem[{S. van~der Walt {et~al.}(2011)van~der Walt, Colbert, \& Varoquaux}]{numpy}
van~der Walt, S., Colbert, S.~C., \& Varoquaux, G. 2011, \bibinfo{title}{The NumPy Array: A Structure for Efficient Numerical Computation,} Computing in Science Engineering, 13, 22, \dodoi{10.1109/MCSE.2011.37}

\bibitem[{H. {van Eerten}(2018){van Eerten}}]{2018IJMPD..2742002V}
{van Eerten}, H. 2018, \bibinfo{title}{{Gamma-ray burst afterglow blast waves},} International Journal of Modern Physics D, 27, 1842002, \dodoi{10.1142/S0218271818420026}

\bibitem[{H. {van Eerten} {et~al.}(2012){van Eerten}, {van der Horst}, \& {MacFadyen}}]{2012ApJ...749...44V}
{van Eerten}, H., {van der Horst}, A., \& {MacFadyen}, A. 2012, \bibinfo{title}{{Gamma-Ray Burst Afterglow Broadband Fitting Based Directly on Hydrodynamics Simulations},} \apj, 749, 44, \dodoi{10.1088/0004-637X/749/1/44}

\bibitem[{H. {van Eerten} {et~al.}(2010){van Eerten}, {Zhang}, \& {MacFadyen}}]{2010ApJ...722..235V}
{van Eerten}, H., {Zhang}, W., \& {MacFadyen}, A. 2010, \bibinfo{title}{{Off-axis Gamma-ray Burst Afterglow Modeling Based on a Two-dimensional Axisymmetric Hydrodynamics Simulation},} \apj, 722, 235, \dodoi{10.1088/0004-637X/722/1/235}

\bibitem[{H.~J. {van Eerten} \& A.~I. {MacFadyen}(2011){van Eerten} \& {MacFadyen}}]{2011ApJ...733L..37V}
{van Eerten}, H.~J., \& {MacFadyen}, A.~I. 2011, \bibinfo{title}{{Synthetic Off-axis Light Curves for Low-energy Gamma-Ray Bursts},} \apjl, 733, L37, \dodoi{10.1088/2041-8205/733/2/L37}

\bibitem[{E. {Waxman} {et~al.}(1998){Waxman}, {Kulkarni}, \& {Frail}}]{1998ApJ...497..288W}
{Waxman}, E., {Kulkarni}, S.~R., \& {Frail}, D.~A. 1998, \bibinfo{title}{{Implications of the Radio Afterglow from the Gamma-Ray Burst of 1997 May 8},} \apj, 497, 288, \dodoi{10.1086/305467}

\bibitem[{E. {Waxman} {et~al.}(2007){Waxman}, {M{\'e}sz{\'a}ros}, \& {Campana}}]{2007ApJ...667..351W}
{Waxman}, E., {M{\'e}sz{\'a}ros}, P., \& {Campana}, S. 2007, \bibinfo{title}{{GRB 060218: A Relativistic Supernova Shock Breakout},} \apj, 667, 351, \dodoi{10.1086/520715}

\bibitem[{P.~K.~G. {Williams} {et~al.}(2017){Williams}, {Clavel}, {Newton}, \& {Ryzhkov}}]{2017ascl.soft04001W}
{Williams}, P. K.~G., {Clavel}, M., {Newton}, E., \& {Ryzhkov}, D. 2017, {pwkit: Astronomical utilities in Python},, Astrophysics Source Code Library, record ascl:1704.001

\bibitem[{S.~E. {Woosley} \& J.~S. {Bloom}(2006){Woosley} \& {Bloom}}]{WoosleyBloom2006}
{Woosley}, S.~E., \& {Bloom}, J.~S. 2006, \bibinfo{title}{{The Supernova Gamma-Ray Burst Connection},} \araa, 44, 507, \dodoi{10.1146/annurev.astro.43.072103.150558}

\bibitem[{E.~L. {Wright} {et~al.}(2010){Wright}, {Eisenhardt}, {Mainzer}, {Ressler}, {Cutri}, {Jarrett}, {Kirkpatrick}, {Padgett}, {McMillan}, {Skrutskie}, {Stanford}, {Cohen}, {Walker}, {Mather}, {Leisawitz}, {Gautier}, {McLean}, {Benford}, {Lonsdale}, {Blain}, {Mendez}, {Irace}, {Duval}, {Liu}, {Royer}, {Heinrichsen}, {Howard}, {Shannon}, {Kendall}, {Walsh}, {Larsen}, {Cardon}, {Schick}, {Schwalm}, {Abid}, {Fabinsky}, {Naes}, \& {Tsai}}]{2010AJ....140.1868W}
{Wright}, E.~L., {Eisenhardt}, P. R.~M., {Mainzer}, A.~K., {et~al.} 2010, \bibinfo{title}{{The Wide-field Infrared Survey Explorer (WISE): Mission Description and Initial On-orbit Performance},} \aj, 140, 1868, \dodoi{10.1088/0004-6256/140/6/1868}

\bibitem[{S.~A. {Yost} {et~al.}(2003){Yost}, {Harrison}, {Sari}, \& {Frail}}]{2003ApJ...597..459Y}
{Yost}, S.~A., {Harrison}, F.~A., {Sari}, R., \& {Frail}, D.~A. 2003, \bibinfo{title}{{A Study of the Afterglows of Four Gamma-Ray Bursts: Constraining the Explosion and Fireball Model},} \apj, 597, 459, \dodoi{10.1086/378288}

\bibitem[{M.~S. {Yun} \& C.~L. {Carilli}(2002){Yun} \& {Carilli}}]{2002ApJ...568...88Y}
{Yun}, M.~S., \& {Carilli}, C.~L. 2002, \bibinfo{title}{{Radio-to-Far-Infrared Spectral Energy Distribution and Photometric Redshifts for Dusty Starburst Galaxies},} \apj, 568, 88, \dodoi{10.1086/338924}

\bibitem[{B.-B. {Zhang} {et~al.}(2012){Zhang}, {Fan}, {Shen}, {Xu}, {Zhang}, {Wei}, {Burrows}, {Zhang}, \& {Gehrels}}]{2012ApJ...756..190Z}
{Zhang}, B.-B., {Fan}, Y.-Z., {Shen}, R.-F., {et~al.} 2012, \bibinfo{title}{{GRB 120422A: A Low-luminosity Gamma-Ray Burst Driven by a Central Engine},} \apj, 756, 190, \dodoi{10.1088/0004-637X/756/2/190}

\end{thebibliography}

\end{document}